\documentclass[12pt]{article}

\usepackage{amssymb}
\usepackage[centertags]{amsmath}
\usepackage{rotating}
\usepackage[dvips]{color}	% For making/using complex colors
\usepackage{epsf}
\usepackage{pazh_eng}

\tightenlines

\newcommand{\m}{^m\!\!\!.\,}

\DeclareMathOperator{\mex}{\mathsf M} %Nii
\DeclareMathOperator{\var}{\mathsf D} %Nii

\sloppypar

\begin{document}

{\renewcommand{\thefootnote}{\fnsymbol{footnote}}

\title{\bf The Axial Zone of Avoidance in the Globular Cluster System
and the Distance to the Galactic Center}

\author{\bf
I. I. Nikiforov\footnote{E-mail: \tt nii@astro.spbu.ru} and E. V. Agladze}

}
\setcounter{footnote}{0}

\vspace{-0.9em}
\begin{center}
{\it Department of Celestial Mechanics, St.\ Petersburg State University, 

\vspace{-.8em}
Universitetskij pr.\ 28, Staryj Peterhof, St.\ Petersburg, 198504 Russia}
\end{center}

\vspace{-0.3em}
%\received{Received May 10, 2016}

\sloppypar 
\vspace{2mm}
\noindent
{\bf Abstract}---We have checked the existence of a zone of avoidance oriented along the
Galactic rotation axis in the globular cluster (GC) system of the Galaxy and performed a
parametrization of this zone in the axisymmetric approximation. The possibility of the
presence of such a structure in the shape of a double cone has previously been discussed
in the literature. We show that an unambiguous conclusion about the existence of an axial
zone of avoidance and its parameters cannot be reached based on the maximization of the
formal cone of avoidance due to the discreteness of the GC system. The ambiguity allows
the construction of the representation of voids in the GC system by a set of
largest-radius meridional cylindrical voids to be overcome. As a result of our structural
study of this set for northern and southern GCs independently, we have managed to identify
ordered, vertically connected axial zones of avoidance with similar characteristics. Our
mapping of the combined axial zone of avoidance in the separate and joint analyses of the
northern and southern voids shows that this structure is traceable at $|Z|\ga 1$~kpc, it
is similar in shape to a double cone whose axis crosses the region of greatest GC number
density, and the southern cavity of the zone has a less regular shape than the northern
one. By modeling the distribution of Galactocentric latitudes for GCs, we have determined
the half-angle of the cone of avoidance $\alpha_0={15\fdg 0}^{+2\fdg1}_{-4\fdg1}$ and the
distance to the Galactic center ${R_0}=7.{\color{black}3}\pm0.{\color{black}5}$~kpc (in
the scale of the Harris (1996) catalog, the 2010 version) as the distance from the Sun to
the point of intersection of the cone axis with the center–anticenter line. A correction
to the calibration of the GC distance scale obtained in the same version of the Harris
catalog from Galactic objects leads to an estimate of ${R_0}=7.{\color{black}2}\pm
0.{\color{black}5}\bigr|_{\text{\color{black}stat}} \pm
0.{\color{black}3}\bigr|_{\text{\color{black}calib}}~\text{kpc}$.   The systematic error
in $R_0$ due to the observational incompleteness of GCs for this method is insignificant.  
The probability that the zone of avoidance at the characteristics found is random in
nature is ${\le}2\%$. We   have revealed evidence for the elongation of the zone of
avoidance in the direction orthogonal to the center--anticenter axis, which, just as the
north--south difference in this zone, may be attributable to the influence   of the
Magellanic Clouds. The detectability of similar zones of avoidance in the GC systems of
external   galaxies is discussed.


\clearpage

%***************************************************************
\section{INTRODUCTION}
\label{introduction}

Globular clusters (GCs) are traditionally used to study the structure, kinematics,
dynamics, chemical evolution, and other properties of the Galactic halo and the Galaxy as
a whole (see, e.g., Borkova and Marsakov 2000; Harris 2001; Ashman and Zepf 2008; Loktin
and Marsakov 2009). In particular, following the pioneering paper by Shapley (1918), the
solar Galactocentric distance ($R_0$) is determined from GCs, with GCs having been the
most popular type of objects for this purpose until recently. As a rule, one or another
modification of Shapley's method was applied to estimate $R_0$ from GCs; it consists in
finding the distance to the centroid of the spatial distribution of GCs, i.e., to the
geometric center or the point of greatest density of the GC system (for recent
implementations see Maciel 1993; Rastorguev et al.\ 1994; Harris 2001; Bica et al.\ 2006).
However, the unfolded discussion of the observational selection of GCs attributable to
extinction and leading to an underestimation of $R_0$ in the classical version of
Shapley's method (see the reviews by Reid (1993), Nikiforov (2003), and the references in
Table 8 of this paper) has stimulated a search for other, apart from a central
concentration, peculiarities of the spatial distribution of GCs that are capable of giving
constraints on~$R_0$ without any significant systematic bias due to the selection effect.
For example, Sasaki and Ishizawa (1978) proposed to use the {\em cone of avoidance\/}
(COA) in the GC system, by which a drop in the number density or a complete absence
of GCs in the double cone whose axis is orthogonal to the Galactic plane and whose vertex
lies at the Galactic center is meant, to determine $R_0$\,.

Wright and Innanen (1972b) were the first to draw attention to the ``apparent dearth of
GCs in a ${\sim}15\deg$ `cone of avoidance' centered on a Galactic nucleus'' in the GC
distribution on the $XZ$ plane by alluding to Fig. 3 in the review of Oort (1965) as an
illustration. Here, $X$ is the axis passing through the Sun and the Galactic center, $Z$
is the axis orthogonal to the Galactic plane, and the Sun is at the coordinate origin.
Wright and Innanen (1972b) assumed a connection between the presence of COA and the
following theoretical result. Wright and Innanen (1972a) investigated the collapse of a
massless, nonrotating gas spheroid around a massive point nucleus without any gas
pressure. They showed analytically that the gas envelope attained an ``infinity-sign''
shape in a meridional section as a result of the collapse (at the instant of the central
singularity). At the same time, the authors pointed out that for a rotating initial
spheroid the result should change insignificantly, in particular, because the angular
momentum per unit mass of a real protogalaxy should be small near the rotation axis.
Wright and Innanen (1972b) associated the deficit of gas arising from the collapse along
the minor axis of the spheroid with the COA in the Galactic GC system. However, the
conclusion about the existence of the COA itself was qualitative, being apparently based
only on the visual impression from the pattern of the GC distribution in projection onto
the $XZ$ plane. The brief article by Wright and Innanen (1972b) contains no mention of any
statistical analysis of the GC distribution, any estimation of its parameters.

Based on the results and ideas of Wright and Innanen (1972a, 1972b) but without touching
on the question of whether the existence of COA in the Galactic GC system was real, Sasaki
and Ishizawa (1978) considered the possible dynamical mechanisms for the formation of this
structure. Sasaki and Ishizawa (1978) numerically simulated the evolution of an initially
spherically symmetric GC system in the Galactic field and showed that the instability of
orbits with a low angular momentum and the tidal disruption of GCs as a result of their
passage near the Galactic nucleus could collectively give rise to COA in the GC system in
a time of ${\sim}10$~Gyr; the peripheral regions of the COA are ``cleaned'' of GCs faster.
In the same paper the COA in the sense of a {\em complete\/} absence of GCs in
it\footnote{Some other name, for example, a ``cone of emptiness'' or a ``GC-free cone,''
would correspond better to this definition. Sasaki and Ishizawa (1978) preferred the
existing term ``cone of avoidance'' introduced by Wright and Innanen (1972b) to designate
the {\em manifestations\/} of this spatial feature when projecting the GC distribution
onto the $XZ$ plane. We retain the latter name for continuity and because it is physically
more correct andmathematically more convenient not to rule out {\em completely\/} the
appearance of a cluster in the ``forbidden zone.'' Wright and Innanen (1972b) introduced
their term apparently by analogy with the term ``zone of avoidance'' for galaxies near the
plane of our Galaxy whose use dates back to R. Proctor (see the review by Kraan-Korteweg
and Lahav (2000)).} was first used to determine the distance to the Galactic center by
which the point of intersection of the COA axis with the $X$ axis was meant. As a result,
Sasaki and Ishizawa (1978) derived an estimate of $R_0 = 9.2 \pm 1.3$~kpc from the COA and
an estimate of the COA half-angle $\alpha_0 = 14\fdg 3 \pm 2\fdg8$.

We have failed to find more recent papers devoted directly to GCs in which the COA would
be investigated or its existence would be questioned or it would simply be mentioned,
although the interest in this structure, if it is real, and its allowance in solving
various problems would seem obvious. At the same time, the $R_0$ estimate obtained by
Sasaki and Ishizawa (1978) from the COA has been traditionally included in reviews on the
problem of $R_0$ determination up until now (de Vaucouleurs 1983; Kerr and Lynden-Bell
1986; Reid 1989, 1993; Surdin 1999; Francis and Anderson 2014); it was taken into account
when deriving the mean (``best'') value $\langle R_0\rangle_{\text{best}}$ in many of its
calculations (de Vaucouleurs 1983; Kerr and Lynden-Bell 1986; Reid 1989, 1993; Surdin
1999; Nikiforov 2004; Nikiforov and Smirnova 2013). In comparison with the present-day
measurements of this parameter, $\langle R_0\rangle_{\text{best}}=(7.8$--$8.25) \pm
(0.1$--$0.5)$~kpc (Reid 1993; Nikiforov 2004; Avedisova 2005; Genzel et al.\ 2010; Foster
and Cooper 2010; Nikiforov and Smirnova 2013; Bland-Hawthorn and Gerhard 2016), the
individual $R_0$ estimates published since 2000 have point values no greater than 8.9 kpc
(Francis and Anderson 2014; Bland-Hawthorn and Gerhard 2016), $R_0 = 9.2$~kpc deduced by
Sasaki and Ishizawa (1978) seems overestimated.

Below we discuss the possible causes of this discrepancy.
However, the main goal of this paper is to
check the very fact of the existence of an axial zone of
avoidance in the Galactic GC system and, in the case
of an affirmative answer to this question, to clarify the
characteristic features of the geometry of this zone
and to redetermine $R_0$ using this spatial structure
from the currently available data on GCs.

It should be noted that the principle of finding $R_0$ from the COA itself proposed by Sasaki
and Ishizawa (1978), irrespective of the systematics of its first realization, can be
promising. Since the position of the COA is determined mainly by the clusters at large $|Z|$
(Sasaki and Ishizawa 1978), i.e., at high $|b|$, the selection effect must have virtually no
influence on the result; the problem of extinction correction when determining the
distances to these GCs is also less acute for them. Therefore, the corresponding
systematic biases of the $R_0$ estimate are expected to be smaller in the case of using the
COA as a key feature of the spatial distribution of GCs than in the case of relying on the
GC centroid in Shapley's method. 
\medskip

\section{DATA ON GLOBULAR CLUSTERS} 

As the database we took the 2010 version of the catalog by Harris (1996) below referred to
as H10. The heliocentric distances calculated using the scale
\begin{equation}
    \label{H10cal}
    M_V(\text{HB})=0.16\,\text{[Fe/H]}+0.84,
\end{equation}
adopted by Harris based on his own calibration from new data are given for all 157
clusters of the catalog. Here, $M_V(\text{HB})$ is the mean absolute $V$ magnitude of the horizontal
branch, and [Fe/H] is the metallicity.

The [Fe/H] estimates given in H10 for 152 GCs were
used in this paper only to separate the metal-rich and metal-poor cluster subsystems. We
adopted the boundary metallicity [Fe/H] = −0.8 (Nikiforov and Smirnova 2013). 
\medskip

\section{THE METHOD OF MAXIMIZING THE FORMAL\\ CONE OF AVOIDANCE} 

The estimate of $R_0=9.2\pm1.3$~kpc deduced by Sasaki and Ishizawa (1978) was obtained for
the catalog of distances with the calibration $M_V({\rm HB})=0{\color{black}\m}5$ that is
brighter than the present-day ones, i.e., gives a longer scale. Rescaling the estimate to
calibration~(1) gives $R_0=8.7\pm1.2$~kpc for $\text{[Fe/H]}=-1.3$, the mean metallicity
for all GCs, and $R_0=8.8\pm1.2$~kpc for  $\text{[Fe/H]}=-1.5$, the mean metallicity for
metal-poor GCs (Nikiforov and Smirnova 2013), which largely determine the COA. Clearly,
these values of~$R_0$ do not contradict the present-day estimates, although they are near
the edge of their range. However, Sasaki and Ishizawa (1978) provided another estimate
from the COA, $R_0=9.4\pm1.2$~kpc, based on one of the early versions of the catalog by
Harris with more accurate GC distances and the calibration $M_V({\rm
HB})=0{\color{black}\m}6$. Rescaling to calibration~(1) barely changes this estimate:
$R_0=9.3\pm1.2$ and~$R_0=9.4\pm1.2$~kpc for $\text{[Fe/H]}=-1.3$  and $-1.5$,
respectively. The latter values suggest that, on the whole, the procedure by Sasaki and
Ishizawa (1978) leads to overestimates of $R_0$ from the COA; the overestimation cannot be
explained by the evolution of the GC distance scale calibration.

Sasaki and Ishizawa (1978) used the natural principle of maximizing the COA opening angle
as the basis for their analysis: the value of $R_0$ at which this angle turns out to be
largest is taken to be the ``true'' one. Obviously, the COA determined in this way in
nondegenerate cases is specified by only two GCs (for an explanation see below, after Eq.
(3)). Sasaki and Ishizawa (1978) did not apply this procedure to the original catalogs of
distances, apparently because the optimal value of $R_0$ in such a case is based only on
two distance estimates having some uncertainties. Instead, they maximized the COA for each
of the 300 pseudo-random GC catalogs obtained by varying the distances moduli of GCs
according to a normal law with a mean corresponding to the cataloged distance and a
standard deviation of ${0\m5}$ or ${0\m3}$  for different original catalogs. The trial
values of $R_0$ were taken in the range from 6 to 12 kpc. The optimal values of $R_0$
found for each generated catalog were then averaged.

To find out what the systematics of this procedure
could be, let us consider the dependence of the largest
(in absolute value) Galactocentric latitude of GCs,
$\varphi_0\equiv\pi/2-\alpha_0= \max
|\varphi|$, on the adopted $R_0$ derived
from the H10 data (Fig. 1a). Here, the Galactocentric
latitude of an individual GC, $\varphi$, at a given~$R_0$ is defined
by the expression
\begin{equation}
\label{phi}
\varphi=\arctan{\frac{Z}{R}}\,,%\nonumber
\end{equation}
where $Z$ is the GC height above the Galactic plane,
and $R$ is the distance from the GC to the Galactic
rotation axis (Galactoaxial distance). In turn,
\begin{equation}
\label{RZ}
Z=r\sin{b}, \qquad
R=\sqrt{R^{2}_{0}+r^2\cos^2b-2R_0r\cos{l}\cos{b}}, %\nonumber
\end{equation}
where $r$ is the heliocentric distance, $l$ and $b$ are the Galactic longitude and
latitude of the GC. Equations (2) and (3) show that the function~$|\varphi|(R_0)$ for an
individual GC has a (single) maximum at minimum $R$ (i.e., at $R_0$ at which the formal
radius vector drawn to the GC is orthogonal to the $X$ axis), while this function has no
minima. Each segment of the function $\varphi_0(R_0)$ in Fig. 1a that has a continuous
derivative for some interval of~$R_0$ is a segment of the dependence~$|\varphi|(R_0)$ for
the GC that has the largest~$|\varphi|$, i.e., determines~$\varphi_0$\,, in this interval
of $R_0$\,. Hence a local minimum of the function~$\varphi_0(R_0)$ can emerge only if at some value
of the argument +$R_0=R_0^*$ the decreasing segment of the function~$|\varphi|(R_0)$ for some GC
determining~$\varphi_0$ at $R_0<R_0^*$ will obtain a value equal to the value of the
increasing segment of the function~$|\varphi|(R_0)$ for another GC determining~$\varphi_0$ at $R_0 >
R_0^*$\,. Thus, the position of each minimum of~$|\varphi|(R_0)$, including the global one, is
determined by the data on only {\em two\/} GCs, i.e., the formal maximization of the COA gives a
solution that eventually ``relies'' on the pair of GCs selected through this procedure.

%xxxxxxxxxxxxxxxxxxxxxxxxxxxxxxxxxxxxxxxxxxxxxxxxxxxxxxxxxxxxxx
\begin{figure}[t!]
%\hspace{-2cm}\epsffile{fi.eps}
%\vspace{-6.2em}
\centerline{%
\epsfxsize=15cm%
\epsffile{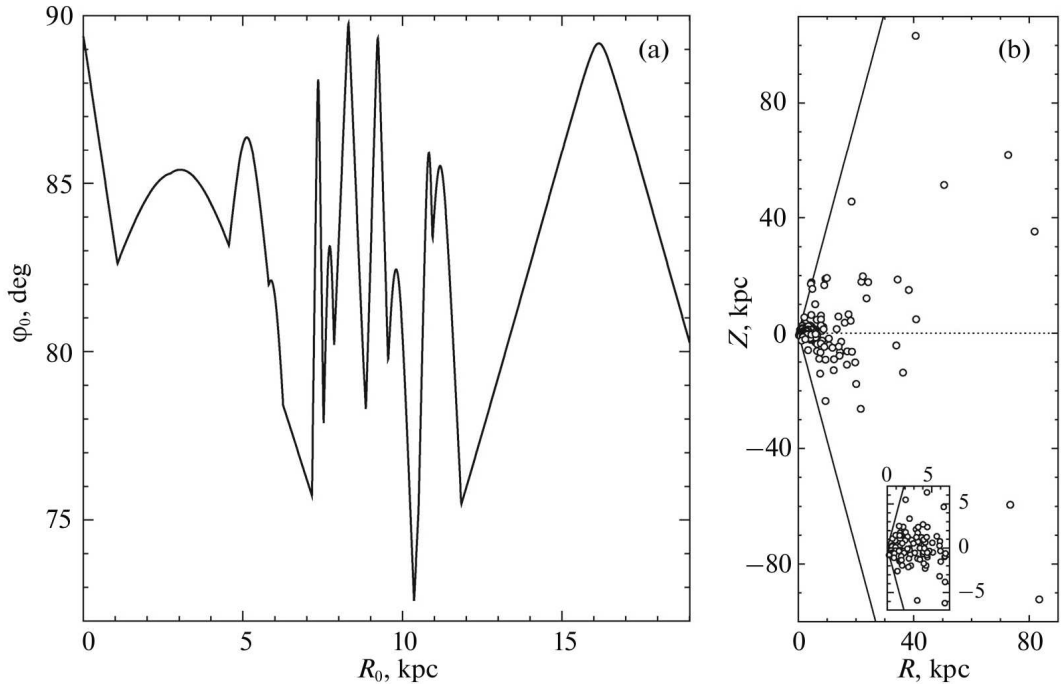}%
%\epsffile{fig1_600.eps}%
}%
%\vspace{-6em}
\caption{\rm (a) Largest (in absolute value) Galactocentric latitude of GCs,
$\varphi_0\equiv\pi/2-\alpha_0$, versus adopted $R_0$\,. (b) The spatial distribution of
GCs in coordinates $(R,Z)$, where $R$ is the Galactoaxial distance, $Z$~is the height above
the Galactic plane, for one of the minima of the dependence~$\varphi_0(R_0)$ ($R_0 = 7.16$~kpc);
the oblique straight lines correspond to the boundaries of the axial cone with a
half-angle $\alpha_0=15^\circ$.} %\label{maxcone}
\end{figure}
%xxxxxxxxxxxxxxxxxxxxxxxxxxxxxxxxxxxxxxxxxxxxxxxxxxxxxxxxxxxxx

The COA must lead to a minimum of depth ${\approx}\alpha_0$ in the plot of the
dependence~$\varphi_0(R_0)$. Figure 1a shows that this dependence actually has three deep
minima (at $R_0 = 7.16$, $10.36$, and $11.85$ kpc) with comparable opening angles
$\alpha_0\ga 15^\circ$ and several shallower ones. Obviously, the numerous local minima
are attributable to the discreteness of the GC system, which increases with distance from
the Galactic center, especially in the post-central region where the undetectability of
GCs increases sharply (Nikiforov and Smirnova 2013). The fact that two of the three
deepest minima occur at $R_0>10$~kpc provides evidence for the latter effect. Strictly
speaking, based only on this dependence, we cannot understand which of the minima
corresponds to the real COA and which are ``spurious.'' The GC distribution on the $RZ$
plane with a characteristic conical region of avoidance along the $Z$~axis shown in
Fig.~1b for the minimum at $R_0 = 7.16$~kpc turns out to be similar for the other two deep
minima as well.

This ambiguity is not removed by varying the distances to clusters. For each pseudo-random
catalog Sasaki and Ishizawa (1978) chose $R_0$ that corresponded to the deepest formal
minimum of~$\varphi_0(R_0)$ without any attempt to distinguish the axial (central) minimum
from the off-axis (peripheral) ones. The depth of the minima also varies when varying the
distances, and if there are two or more comparable minima, then sometimes one, sometimes
another can become the deepest of them. Consequently, averaging the formal solutions over
all model catalogs either is actually averaging the values of $R_0$ for the deepest minima
of the initial dependence~$\varphi_0(R_0)$ or gives the average position of one (deepest)
minimum, which may turn out to be an off-axis one. In the first case, the mean $R_0$ is
found to be, in general, shifted, while its variance more likely reflects the scatter in
the positions of the deep minima than the influence of the errors in the distances on the
position of the deepest of them. Indeed, a formal averaging of the positions of three
minima in Fig.~1a gives $R_0=9.8\pm 1.4$~kpc, a value close to the point estimates and
uncertainties in the results from Sasaki and Ishizawa (1978). Even if the second case is
realized, the global minimum of~$\varphi_0(R_0)$ will very likely to be an off-axis one.
For the original H10 data it turns out to be such, leading to an even larger $R_0 =
10.4$~kpc (Fig.~1a). Note that the dependences~$\varphi_0(R_0)$ constructed by us from
earlier GC catalogs for various samples are similar in basic features to the dependence in
Fig.~1a: there were a total of two or three comparable deep minima the deepest of which,
as a rule, was at $R_0\sim 10$~kpc; the structure of the middle part of the dependence
only became more complicated with time due to the appearance of a larger number of central
GCs in the catalogs. Sasaki and Ishizawa (1978) did not provide the
curve~$\varphi_0(R_0)$.

Since the discreteness of the GC system behind the Galactic center is higher because of
the selection effect (Nikiforov and Smirnova 2013), the occurrence probability of deep
off-axis minima is also higher there (Fig.~1a). This explains why the values obtained by
Sasaki and Ishizawa (1978) were overestimated.

Thus, applying the method of maximizing the formal COA generally leads to an incorrect
result, irrespective of whether the GC distances are varied or not. Modeling the
distribution of Galactocentric latitudes for GCs whereby the $R_0$ estimate is determined
by the spatial distribution of {\em all\/} GCs from the sample can be an alternative. The
dependences $|\varphi|(R_0)$ only for {\em two\/} GCs are eventually used in maximizing
the COA.

However, before we turn to this modeling, the very fact of the existence of a region of
avoidance should be checked, because, strictly speaking, the mere presence of deep minima
in the plot of~$\varphi_0(R_0)$ does not prove this fact: all these minima can be off-axis
ones attributable to the discreteness of the GC system. If the existence of a region of
avoidance will be confirmed, then it is also necessary to establish the main geometric
properties of this region.
\medskip

\section{MAPPING THE AXIAL ZONE OF AVOIDANCE\\ IN THE GLOBULAR CLUSTER SYSTEM}

%.............................................................
\subsection{Representation of Voids by a Set of Meridional Cylindrical Voids}
\label{CVs}

To study the regions of avoidance in the GC system oriented along the $Z$~axis, let us
partition the entire space into layers whose boundaries are parallel to the Galactic
plane. We will begin the partition in the northward and southward directions from the
Galactic plane ($Z = 0$~kpc) and will select the boundaries of the layers in such a way
that the number of GCs in each layer is no smaller than ten. As a result of the partition,
we obtained 14~$Z$~layers and, accordingly, GC samples. Following the principle of
maximizing the region of avoidance (Sasaki and Ishizawa 1978), we will search for the
largest voids in each layer along the local (passing through the Sun) meridional plane of
the Galaxy. For this purpose, we constructed the dependence ${\rho}(X)$ for each
$Z$~layer, where $\rho$ is the distance from a point on the $X$ axis ($X$ is the
coordinate of the point) to the nearest GC projection onto the $XY$ plane (Galactic
plane); the $Y$ axis points in the direction of Galactic rotation. In this dependence we
found local maxima ${\rho}_{\text{c}}\equiv {\rho}(X_{\text{c}})$, each of which specifies
the locally largest (in volume) cylindrical region of avoidance (in which there are no
GCs) with radius~${\rho}_{\text{c}}$ and axis coordinate~$X_{\text{c}}$ (here, ``c''
stands for cylinder). Below we will call such regions {\em cylindrical voids\/} (CVs).
Here, we restrict our analysis only to the {\em meridional\/} CVs whose axes lie in the
local meridional plane of the Galaxy. A set of CVs was found for each $Z$~layer. As an
example, Fig.~2 presents the dependences ${\rho}_{\text{c}}(X)$ for the southern
$Z$~layers, while Fig.~3 shows the projections of the CVs found for these layers onto the
$XY$ plane. Also, the open and filled circles of the same size in Fig.~3 indicate,
respectively, the GCs determining the CVs and the remaining GCs that fell into the
$Z$~layer; the small filled circles mark the positions of the CV axes.

%xxxxxxxxxxxxxxxxxxxxxxxxxxxxxxxxxxxxxxxxxxxxxxxxxxxxxxxxxxxxxx
\begin{figure}[t!]
%\hspace{-2cm}\epsffile{fi.eps}
%\vspace{-6.2em}
\centerline{%
\epsfxsize=15cm%
\epsffile{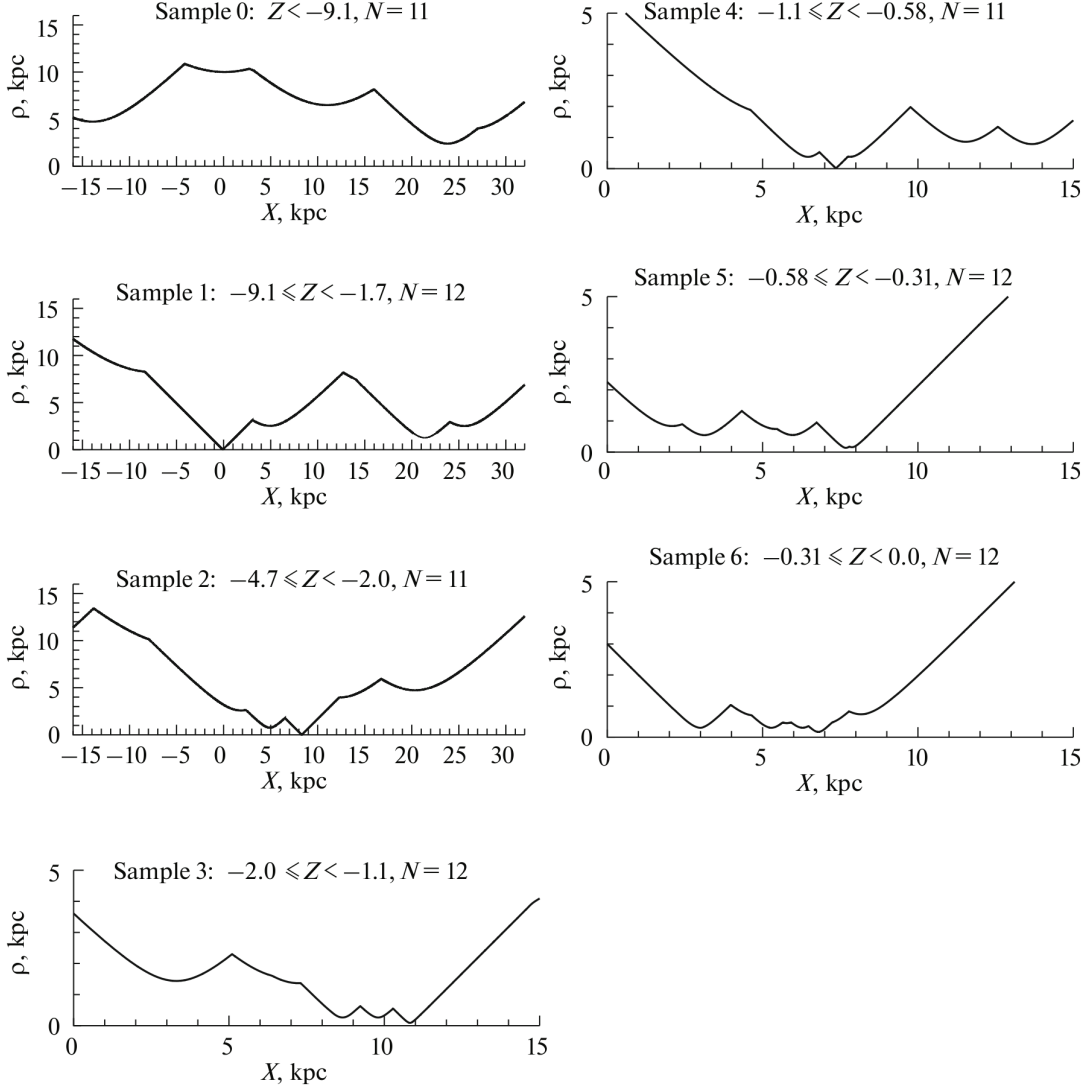}%
}%
%\vspace{-6em}
\caption{\rm Distance $\rho$ from a point on the $X$ axis to the nearest GC projection onto the
Galactic plane versus coordinate $X$ of this point for the southern $Z$~layers. The origin of
the $X$ axis corresponds to the position of the Sun.}
%\label{maxcone}
\end{figure}
%xxxxxxxxxxxxxxxxxxxxxxxxxxxxxxxxxxxxxxxxxxxxxxxxxxxxxxxxxxxxx

%xxxxxxxxxxxxxxxxxxxxxxxxxxxxxxxxxxxxxxxxxxxxxxxxxxxxxxxxxxxxxx
\begin{figure}[t!]
%\hspace{-2cm}\epsffile{fi.eps}
%\vspace{-6.2em}
\centerline{%
\epsfxsize=15cm%
\epsffile{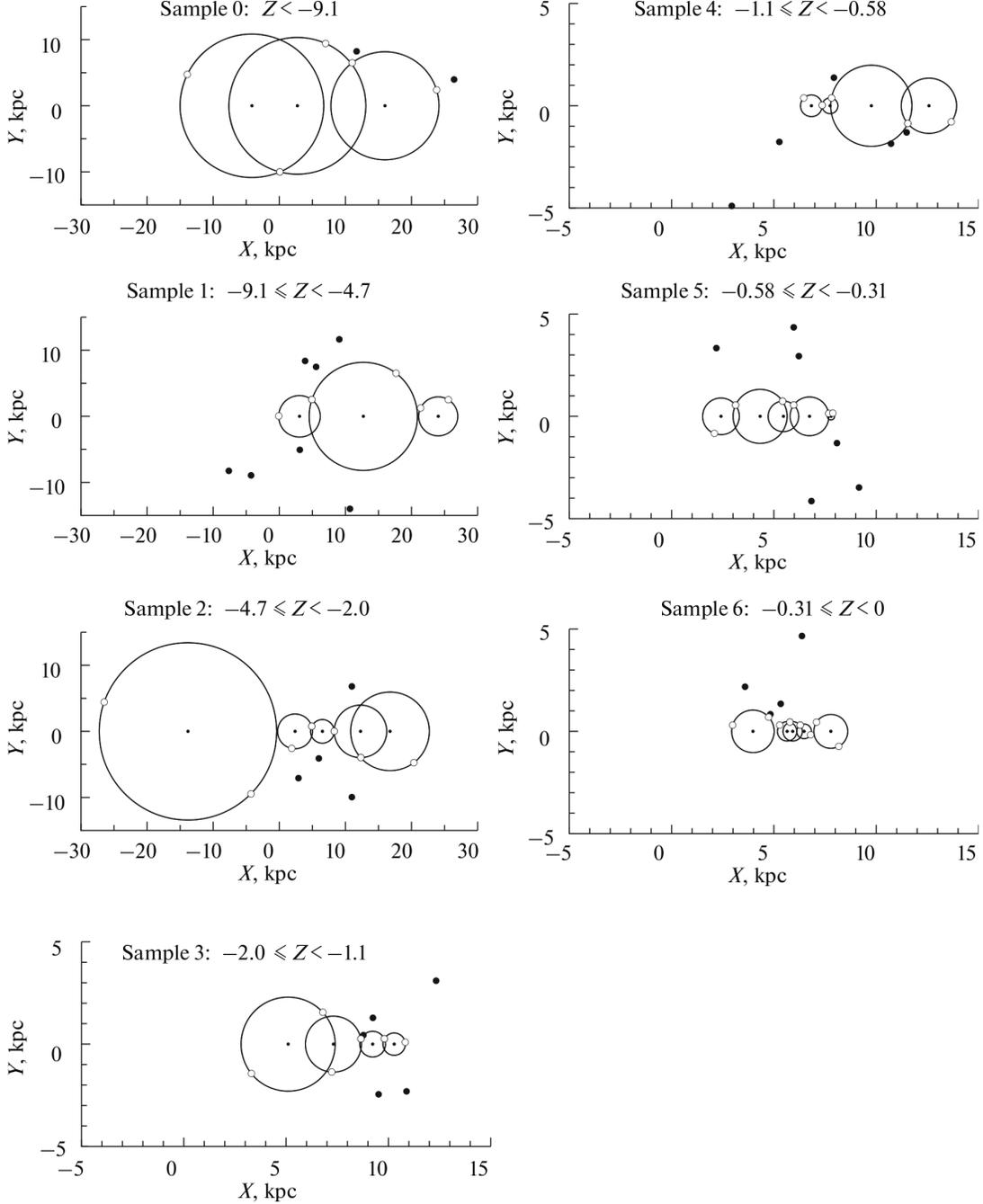}%
}%
%\vspace{-6em}
\caption{\rm 
Cylindrical voids found in projection onto the Galactic plane ($XY$) for the southern $Z$
layers. The open circles are the GCs determining the voids; the large filled circles are
the remaining clusters falling into a $Z$~layer; the small filled circles are the positions
of the void axes. The Sun is at point $(X,Y)=(0,0)$.}
%\label{maxcone}
\end{figure}
%xxxxxxxxxxxxxxxxxxxxxxxxxxxxxxxxxxxxxxxxxxxxxxxxxxxxxxxxxxxxx

The projection of the constructed CV system onto the $XZ$ plane allows the void
distribution pattern as a whole to be imagined. Let us introduce new terms and
designations. As can be seen from Fig.~3, CVs can intersect. If (two or more) voids
intersect, i.e., if they are {\em nonisolated\/}, then the single {\em combined void\/}
formed by them is considered. The vertical boundaries of the latter in projection onto the
$XZ$ plane are the leftmost and rightmost boundaries of the CVs constituting it; they will
be depicted by solid lines. We will designate the edges of the CV intersections by long
dashes and the boundaries of the nonisolated CVs that fell into the combined voids by
short dashes. Figure 4 (lower panel) gives an example of the projection of one of the
$Z$~layers onto the $XZ$ plane. The upper and middle panels of Fig.~4 for the same
$Z$~layer show the dependence ${\rho}(X)$ and the CV projections onto the $XY$ plane,
respectively. On the two lower panels of Fig.~4, just as in Fig.~3, the open and filled
circles depict, respectively, the GCs determining the CVs and the remaining GCs that fell
into the $Z$~layer.

%xxxxxxxxxxxxxxxxxxxxxxxxxxxxxxxxxxxxxxxxxxxxxxxxxxxxxxxxxxxxxx
\begin{figure}[t!]
%\hspace{-2cm}\epsffile{fi.eps}
%\vspace{-6.2em}
\centerline{%
\epsfxsize=8cm%
\epsffile{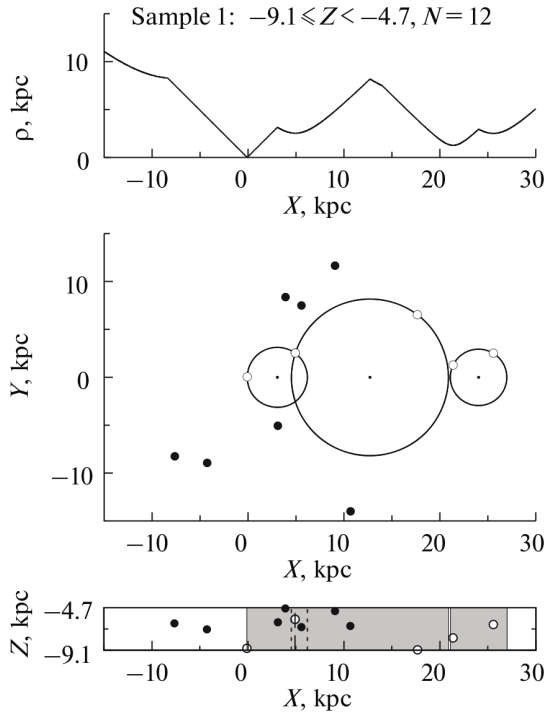}%
}%
%\vspace{-6em}
\caption{\rm 
Finding the locally maximal CVs and constructing their scheme for one of the $Z$~layers as
an example. The upper panel: the function ${\rho}(X)$ whose local maxima,
${\rho}_{\text{c}}\equiv {\rho}(X_{\text{c}})$, determine the axis coordinates ($X_{\text{c}}$) and
radii (${\rho}_{\text{c}}$) of the voids. The middle panel: the projection of the voids onto the $XY$
plane. The lower panel: the projection of the $Z$ layer onto the $XZ$ plane. The gray
shading depicts the void cavities. The designations for GCs are the same as those in
Fig.~3.} %\label{maxcone}
\end{figure}
%xxxxxxxxxxxxxxxxxxxxxxxxxxxxxxxxxxxxxxxxxxxxxxxxxxxxxxxxxxxxx

Using the introduced designations, let us construct a general scheme of the CV
distribution in space. Figure 5 shows the projection of the entire set of CVs identified
in 14 $Z$ layers onto the $XZ$ plane. Since the height of the $Z$ layers increases
dramatically with $|Z|$ due to the rapid drop in the number density of clusters, we will
use a logarithmic scale for the $Z$ coordinate in this and subsequent figures presenting
the zone of avoidance: $\pm \log(1 + |Z|)$. For comparison, Fig.~5 plots the contours of
the presumed COAs (the curves in the adopted coordinates) with half-angles
$\alpha_0=14^\circ$ and $18^\circ$ corresponding to the scatter of estimates for this
parameter in Sasaki and Ishizawa (1978); here, $R_0 = 7.9$~kpc (Nikiforov 2004; Nikiforov
and Smirnova 2013).

%xxxxxxxxxxxxxxxxxxxxxxxxxxxxxxxxxxxxxxxxxxxxxxxxxxxxxxxxxxxxxx
\begin{figure}[t!]
%\hspace{-2cm}\epsffile{fi.eps}
%\vspace{-6.2em}
\centerline{%
\epsfxsize=15cm%
\epsffile{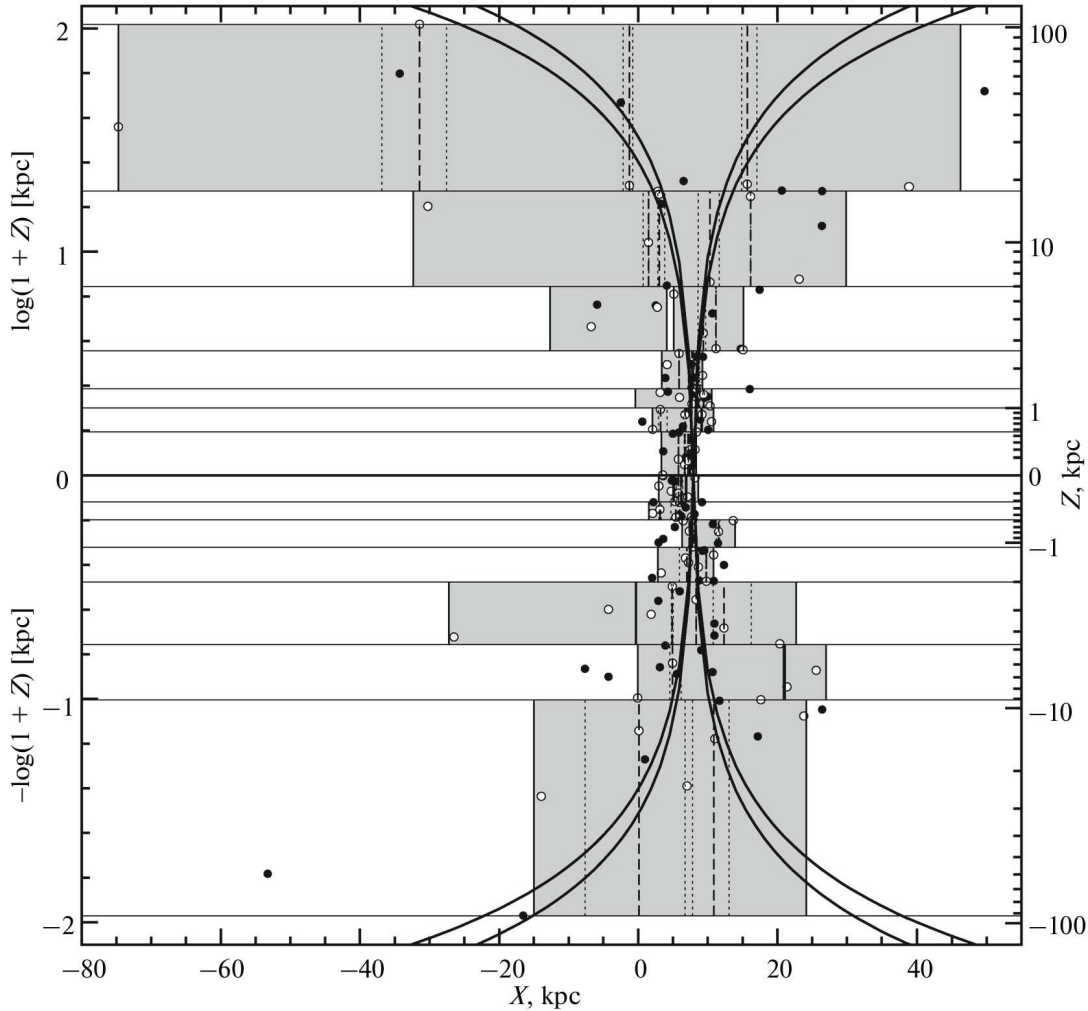}%
}%
%\vspace{-6em}
\caption{\rm
Scheme of the entire set of CVs identified in the GC system. The void cavities are
highlighted by the gray shading. The vertical lines have the same meaning as those on the
lower panel of Fig.~4 (see the text). The curves depict the contours of the axial cones
with $\alpha_0=14^\circ$  and $18^\circ$ for $R_0 = 7.9$~kpc. The designations for GCs are the same as
those in Fig.~3.}
%\label{maxcone}
\end{figure}
%xxxxxxxxxxxxxxxxxxxxxxxxxxxxxxxxxxxxxxxxxxxxxxxxxxxxxxxxxxxxx
\medskip

%.............................................................
\subsection{Identification of the Axial Zone of Avoidance
through a Separate Analysis\\ of the Northern and
Southern Voids}
\label{finding_semi}

Figure 5 shows that the voids form a cone-shaped structure, but wider than the presumed
COA. The latter is not surprising, because many of the CVs found are, obviously, only
formal constructions and do not belong to the sought-for {\em axial\/} structure due to
the discreteness of the GC distribution. Therefore, from the entire set of CVs we will
separate out the subsets of voids each of which forms a {\em vertically connected\/}
structure. As a criterion for constructing such structures and a subset of CVs we will
take the existence of an interval (with zero length inclusive) on the $X$ axis in which
all of the rays with the origins on this axis directed along the $Z$ axis in the northward
or southward direction pass, respectively, all northern or southern $Z$ layers only
through the cavities of the voids of this subset without anywhere crossing the region of
the $Z$ layer where there is no void. We will call this algorithm a {\em semi-through\/}
one. Thus, in this step we consider the northern and southern voids separately. If the
axial zone of avoidance in the GC system is real, then it must be detected in the northern
and southern halves of the Galaxy independently and with similar characteristics.

As a result of the above analysis, we found five northern and three southern vertically
connected subsets of CVs. Let us identify those CVs that are {\em common\/} to all
northern or all southern subsets. Obviously, we can talk about the existence of an axial
zone of avoidance only if such common CVs will be found and will constitute an ordered
structure.

The scheme of the revealed common voids (Fig.~6) shows that both take place. The common
CVs form a generally ordered axial zone of avoidance similar in shape to a double conical
surface. Below we will call these voids {\em axial\/}. As can be seen from Fig.~6, no
significant deviations from the conical shape are observed; the dispersion of the void
boundaries relative to this model, but without any obvious systematic bias, is higher only
in the southern part of the Galaxy. There is no evidence for any serious inclination of
the axis of the zone of avoidance to the Galactic plane either; otherwise we would obtain
a noticeably asymmetric picture of the void boundaries. It is obvious that, for example,
the cylindrical model for the zone of avoidance is unsuitable, because the radii of the
axial CVs, on the whole, increase with $|Z|$ (Fig.~6).

%xxxxxxxxxxxxxxxxxxxxxxxxxxxxxxxxxxxxxxxxxxxxxxxxxxxxxxxxxxxxxx
\begin{figure}[t!]
%\hspace{-2cm}\epsffile{fi.eps}
%\vspace{-6.2em}
\centerline{%
\epsfxsize=15cm%
\epsffile{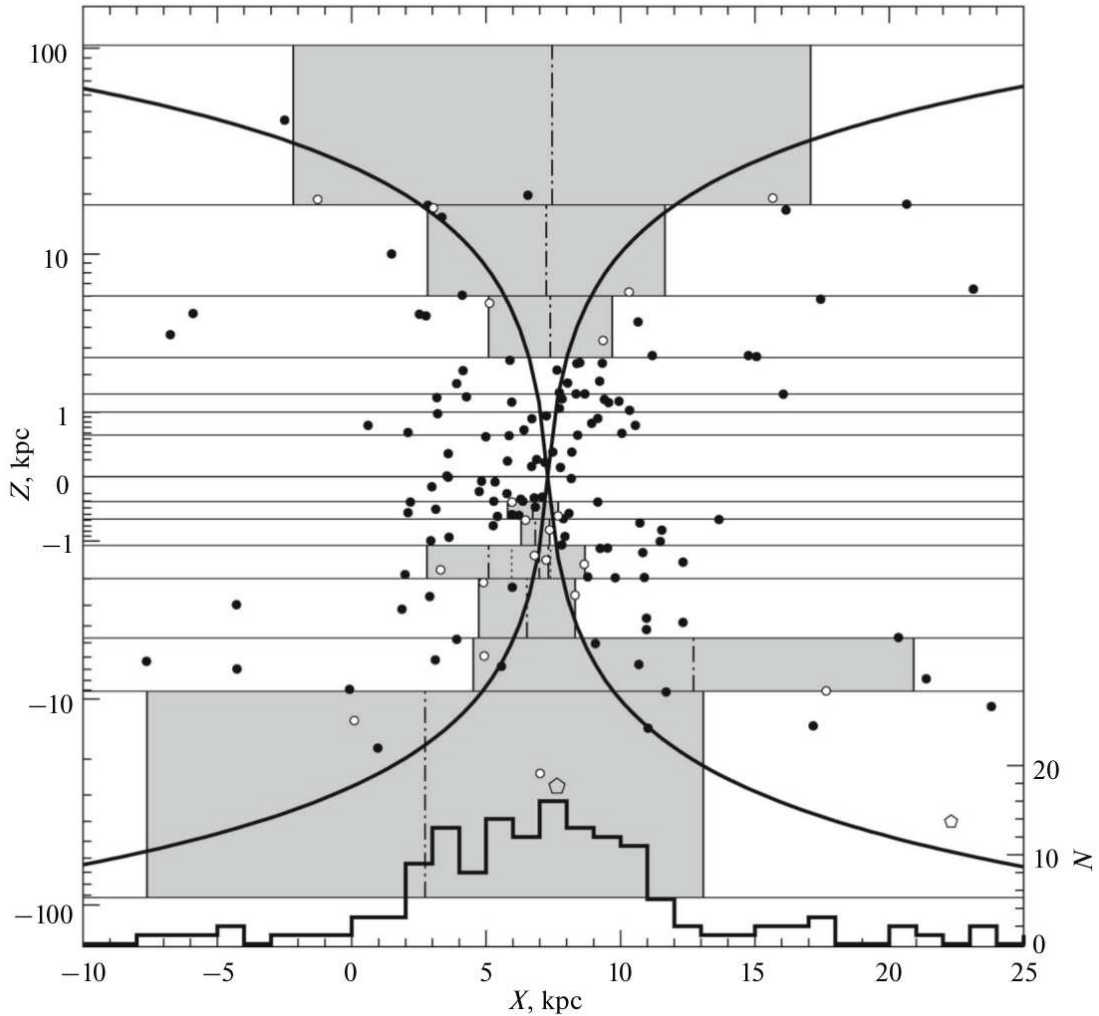}%
%\epsffile{fig6.eps}%
}%
%\vspace{-6em}
\caption{\rm
Axial zone of avoidance in the GC system from the results of our separate analysis of the
southern and northern voids. The shading designates the axial voids; the dash--dotted
lines indicate their axes. The solid curves mark the contour of the axial cone with
$\alpha_0=15\fdg 0$ for $R_0 = 7.3$~kpc (see Subsection 5.3). The histogram is the GC
distribution along the $X$ axis. The pentagons are the LMC and SMC galaxies. The remaining
designations are the same as those in Fig.~5.} %\label{maxcone}
\end{figure}
%xxxxxxxxxxxxxxxxxxxxxxxxxxxxxxxxxxxxxxxxxxxxxxxxxxxxxxxxxxxxx

Comparison with the GC distribution along the $X$ axis (the histogram in Fig.~6) shows
that the positions of the axes of both northern and southern parts of the revealed zone of
avoidance are close to the maximum of the GC number density at $X\approx 7$--$8$~kpc. The
opening angle of the zone of avoidance roughly corresponds to expectations (Fig.~6).

The axial structure is not unambiguously identified in the layers at small $|Z|$. This is
consistent with the results of numerical experiments by Sasaki and Ishizawa (1978),
showing that the COA is ``cleaned'' of GCs primarily in the peripheral regions, while near
the Galactic center GCs must be preserved within the formal COA even on time scales
of~${\sim}20$~Gyr.
\looseness=-1

The axis coordinates ($X_{\text{c}}$) of the identified axial CVs were processed as a
series of equally accurate measurements and as a series of measurements with weights
$p=1/\rho_{\text{c}}$\,, where $\rho_{\text{c}}$ is the CV radius. The choice between
these two cases is ambiguous, because, on the one hand, $X_{\text{c}}$ is determined more
accurately at a small void radius and, on the other hand, the axial zone itself is mainly
determined by the voids at larger $|Z|$, where the void radii are also larger. We
considered the entire set of axial voids as well the northern and southern voids
separately. The processing results are presented in Tables 1 and 2. The notation: $N$ is
the number of voids, $\overline{X_{\text{c}}}$ is the mean value of $X_{\text{c}}$,
${\sigma}_{X_{\text{c}}}$ is the standard deviation of $X_{\text{c}}$ from
$\overline{X_{\text{c}}}$, $\sigma_{0,X_{\text{c}}} $ is the mean error per unit weight,
and ${{\sigma^{\prime}}}_{{X}_{\text{c}}}$ is the weighted standard deviation of
$X_{\text{c}}$ from $\overline{X_{\text{c}}}$\,. For the adopted mean position of the axis
$\overline{X_{\text{c}}}$ in each axial void the formal COA is specified by one of the two
void-forming GCs that has the largest absolute value of the Galactocentric latitude,
$\varphi_{\text{m}}\equiv \max (|\varphi_1|,|\varphi_2|)$. Tables 1 and 2 give the
statistics $\overline{\varphi_{\text{m}}}$, ${\sigma}_{\varphi_{\text{m}}}$,
and~$\varphi_0$\,, i.e., the arithmetic mean, standard deviation, and maximum value of
$\varphi_{\text{m}}$\,, respectively.
\looseness=-1

Tables 1 and 2 show that the values of $\overline{X_{\text{c}}}$ for all samples and
averagings differ only within the error limits. At the same time,
$\overline{X_{\text{c}}}\approx 7$ kpc in all cases, except one case (Table~1,
$Z<0^\circ$) for which the uncertainty in $\overline{X_{\text{c}}}$ is great. The values
of $\overline{X_{\text{c}}}\approx 7$~kpc are close to one of the minima in Fig.~1a ($R_0
= 7.16$~kpc); it apparently corresponds, to a first approximation, to the position of the
COA axis; the remaining minima in this figure are off-axis ones. The values of~$\varphi_0$
for GCs also turn out to be similar.

There is a north--south difference in $\overline{\varphi_{\text{m}}}$\,. There are more
void-forming GCs with relatively small $|\varphi|$ in the southern part of the axial zone
of avoidance; besides, in the layer $-2.0\text{ kpc} \le Z < -1.1$~kpc, i.e., relatively
close to the Galactic plane, we found two nonisolated axial voids forming an extended
structure along the $X$ axis (Fig.~6). This is responsible for the smaller
$\overline{\varphi_{\text{m}}}$ for $Z < 0$~kpc. For the northern part of the zone the
uncertainty in $\overline{X_{\text{c}}}$, along with all other dispersion characteristics,
turn out to be noticeably smaller than those for the southern one (Tables 1 and 2), which
reflects a more regular pattern of the northern cavity of the zone of avoidance (Fig.~6).

On the whole, the results obtained argue for the existence of a zone of avoidance for GCs
similar in shape to a double cone along the Galactic axis (outside the small central
region). The northern and southern cavities of the COA manifest themselves independently
and with similar parameters ($\overline{X_{\text{c}}}$, $\varphi_0$). The latter gives us
grounds to perform a joint analysis of the northern and southern CVs in the next
subsection to increase the identifiability of axial voids.
\medskip

%.............................................................
\subsection{Identification of the Axial Zone of Avoidance by
Analyzing the Entire Set of Voids}
\label{finding_full}

Let us return to the complete set of CVs constructed in Subsection 4.1. Let us separate
out the subsets of voids from it each of which forms a {\em through\/} vertically
connected structure. Thus, in contrast to the semi-through algorithm of Subsection 4.2, as
a criterion for the separation of a CV subset we will take the existence of an interval
(with zero length inclusive) on the $X$ axis in which all of the straight lines crossing
this axis and lying parallel to the $Z$ axis pass all northern {\em and\/} southern $Z$
layers only through the cavities of the voids of this subset without anywhere crossing the
region of the Z layer where there is no void. We will call this algorithm a {\em
through\/} one.

As a result of this analysis, we found 22 different combinations of voids constituting
through vertically connected structures. The identification of CVs common to all these
combinations gives eight axial voids (four northern and four southern ones). The scheme of
the latter is presented in Fig.~7; the characteristics of the axial zone formed by them
are given in Tables 3 and 4, where the designations are the same as those in Tables~1
and~2.

%xxxxxxxxxxxxxxxxxxxxxxxxxxxxxxxxxxxxxxxxxxxxxxxxxxxxxxxxxxxxxx
\begin{figure}[t!]
%\hspace{-2cm}\epsffile{fi.eps}
%\vspace{-6.2em}
\centerline{%
\epsfxsize=15cm%
\epsffile{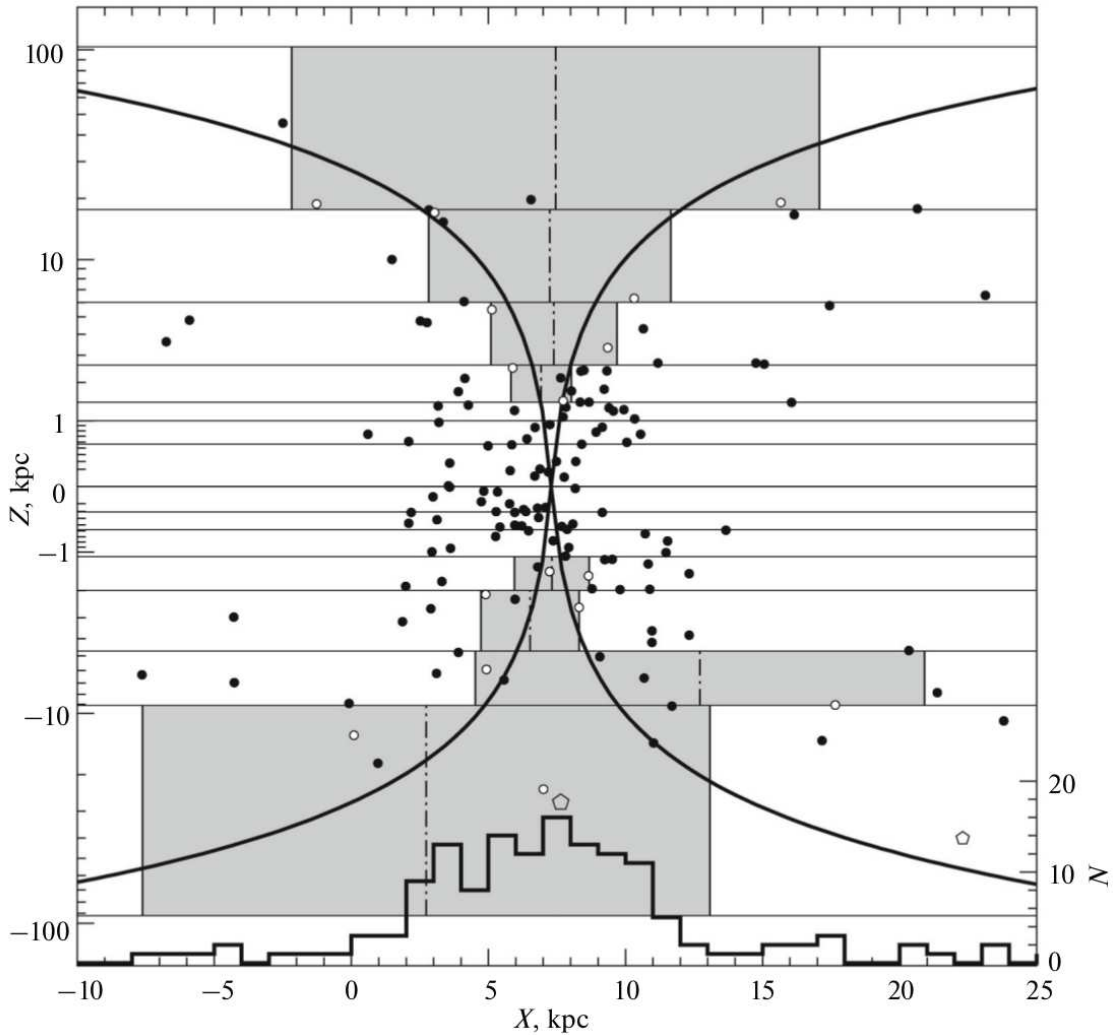}%
}%
%\vspace{-6em}
\caption{\rm
Axial zone of avoidance in the GC system from the results of our analysis of the entire
set of voids. The designations are the same as those in Fig.~6.}
%\label{maxcone}
\end{figure}
%xxxxxxxxxxxxxxxxxxxxxxxxxxxxxxxxxxxxxxxxxxxxxxxxxxxxxxxxxxxxx

%xxxxxxxxxxxxxxxxxxxxxxxxxxxxxxxxxxxxxxxxxxxxxxxxxxxxxxxxxxxxxx
\begin{figure}[t!]
%\hspace{-2cm}\epsffile{fi.eps}
%\vspace{-6.2em}
\centerline{%
\epsfxsize=15cm%
\epsffile{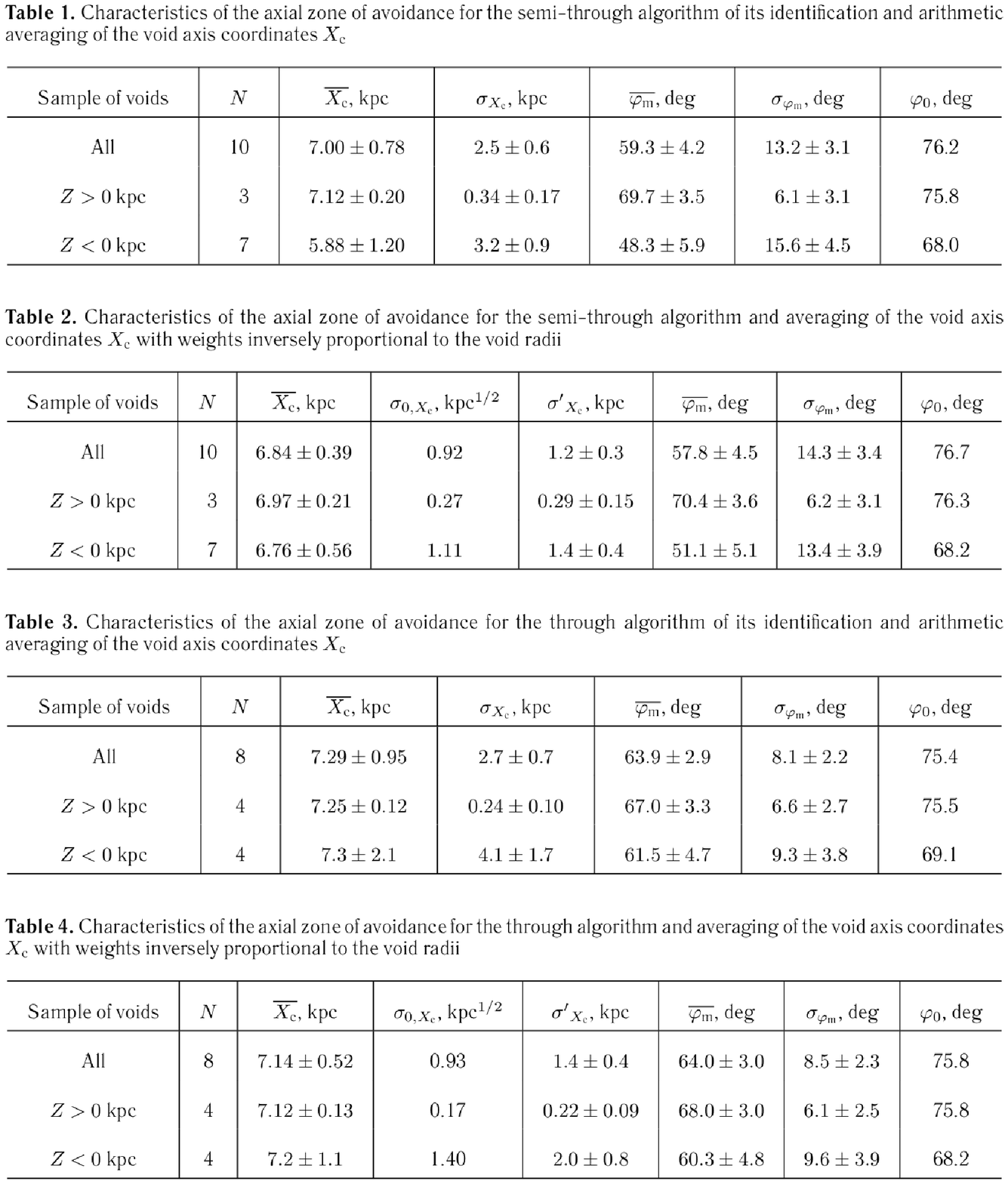}%
}%
%\vspace{-6em}
%\label{maxcone}
\end{figure}
%xxxxxxxxxxxxxxxxxxxxxxxxxxxxxxxxxxxxxxxxxxxxxxxxxxxxxxxxxxxxx

The new results turned out to be generally similar to those based on the semi-through
algorithm. However, the zone of avoidance constructed using the through algorithm is more
symmetric, both relative to the Galactic plane and relative to the mean axis, and more
ordered (cf. Figs. 6 and 7). In particular, this manifests itself in the facts that, in
contrast to the results of the previous subsection, the number of northern and southern
axial voids turned out to be the same, there are no nonisolated CVs among them, the
$Z$~boundaries of unambiguous identification of the axial zone closest to the Galactic
plane are similar in absolute value for the northern and southern cavities of the zone,
respectively, $Z_{\text{lim}}=+1.435$ и $-1.1$~kpc ($Z_{\text{lim}}=+2.6$  and
$-0.31$~kpc were obtained in Subsection 4.2), the north--south difference in
$\overline{\varphi_{\text{m}}}$  and ${\sigma}_{\varphi_{\text{m}}}$ decreased
dramatically, the difference between the formal angular boundaries of~$\varphi_0$ for the
northern and southern COAs was slightly reduced. In addition, the positions of the mean
axis of the zone of avoidance in {\em all\/} cases turned out to be very close
($7.12\leq\overline{X_{\text{c}}}\leq 7.3$~kpc), the scatter of
$\overline{\varphi_{\text{m}}}$ in different cases became much smaller
($60^\circ$--$68^\circ$ versus $48^\circ$--$70^\circ$ in Subsection 4.2), the standard
deviation~${\sigma}_{\varphi_{\text{m}}}$ was generally also reduced (cf. Tables 3, 4 and
Tables 1, 2).

Thus, the through algorithm identifies a zone of avoidance with a more regular structure
closer to a conical one. Although the main north--south differences noted in the previous
subsection turn out to be smaller, they do not disappear completely. They apparently
reflect the objective properties of the spatial distribution of Galactic GCs.

Note that in all cases in Tables 3 and 4 the estimates of $\overline{\varphi_{\text{m}}}$
and ${\sigma}_{\varphi_{\text{m}}}$ differ significantly from the mean
$\mex\Phi_{\text{m}}=\pi/4=45\deg$ and standard deviation
$(\var\Phi_{\text{m}})^{1/2}=\frac{\sqrt{\pi^2-8}}{4}\approx 19\fdg6$, respectively, for a
spherically symmetric GC distribution without COA (see Appendix A2). This suggests that
the choice of voidforming clusters using the through algorithm is not random and argues
for the reality of the voids in the distribution identified by them.

Our results of mapping the zone of avoidance in the Galactic GC system confirm the
existence of an axial COA with a half-angle $\alpha_0=\pi/2-\varphi_0\approx
13^\circ$--$15^\circ$ for the system as a whole. Note that for all samples of axial voids
the formal COA boundary~$\varphi_0$ turned out to be stable with respect to the algorithm
of identifying the zone of avoidance and the method of processing the void axis
coordinates $X_{\text{c}}$ (Tables 1--4).
\medskip

%*************************************************************
\section{MODELING THE DISTRIBUTION OF
GALACTOCENTRIC LATITUDES FOR
GLOBULAR CLUSTERS}
\label{modelling}

Having confirmed the existence of COA in the GC system, let us return to the problem of
using this structure to determine $R_0$ under the assumption of an axial symmetry of the
GC system by taking the point of intersection of the COA axis with the center--anticenter
line as the Galactic center. To overcome the shortcomings of the method of maximizing the
formal COA (see Section 3), we will seek $R_0$ as a parameter of the model distribution
function of Galactocentric latitudes~$\varphi$ for GCs. In this approach the $R_0$
estimate will be a function of the positions of all clusters from the sample rather than
two of them, as in the COA maximization method.

In this case, the results of Section 4.3 may be used
as a priori information, but we may dispense with this.
The first and second methods suggest modeling the $\varphi$
distribution only for the void-forming GCs and for the
complete GC sample without additional data on the
COA, respectively. Each method has its advantages
and disadvantages. Consider both cases.
\medskip
\pagebreak[4]

%.............................................................
\subsection{Modeling for the Void-Forming Globular
Clusters of the Axial Zone of Avoidance}
\label{modelling_voidmarkers}

For the objects that are exactly on the COA surface the minimum (zero value) of the
variance of the absolute values of their Galactocentric latitudes,
${\sigma}^2_{|\varphi|}$, is reached at a trial $R_0$ equal to the true one. If the
void-forming GCs identified in Section 4.3 are assumed to be located near the COA surface,
then minimizing ${\sigma}^2_{|\varphi|}$ for these GCs gives an estimate of~$R_0$ as a COA
parameter. We will then consider the conditional equations written for~$N$~void-forming
clusters 
\begin{equation}
\label{system}
{{|\varphi}_i}(R_0)|=\overline{|\varphi|} ,\quad i=1,\:\ldots,\:N,
\end{equation}
where $\overline{|\varphi|}$ is the mean value of $|\varphi|$ at given $R_0$\,, and the
function $\varphi(R_0)$ is defined by Eqs.~(2) and (3), as an overdetermined system of
equations and will solve it by the least-squares method for the unknowns $R_0$ and
$\overline{|\varphi|}$. Here, we actually assume that $|\varphi|$ for these clusters are
distributed according to the normal law ${\cal
N}(\overline{|\varphi|},{\sigma}^2_{|\varphi|})$, $\overline{|\varphi|}<\varphi_0$\,.

The results of solving system (4) for the complete sample of void-forming GCs as well as
for the northern and southern subsamples of these objects are presented in Table~5. The
standard errors are given for $\overline{|\varphi|}$ and ${\sigma}_{|\varphi|}$. In the
case of a nonlinear parameter $R_0$\,, we took the projection of the two-dimensional
confidence region with a~$1\sigma$ confidence level onto the $R_0$ axis as the confidence
interval (Press et al.\ 1997). The boundaries of this interval were found as the roots of
the equation 
\begin{equation} 
\label{vs12am=}
    \varsigma^2(R_0)=\varsigma^2_0\left(1+{\frac{1}{N_{\text{free}}}}\right),
\end{equation} 
where
\begin{equation} 
\label{vs0212}
    \varsigma^2_0\equiv \min\left[{\sigma}^2_{|\varphi|}\left(R_0,\overline{|\varphi|}\right)\right],
    \qquad
    \varsigma^2(R_0)\equiv \min_{R_0\,=\:\text{const}}
	\left[{\sigma}^2_{|\varphi|}\left(R_0,\overline{|\varphi|}\right)\right],
\end{equation} 
$N_{\text{free}}$ is the number of degrees of freedom (Nikiforov
1999, 2003; Nikiforov and Kazakevich 2009).
We will call the dependence $\varsigma^2(R_0)$ the {\em profile of the
objective function for the parameter $R_0$\/}\,.

%xxxxxxxxxxxxxxxxxxxxxxxxxxxxxxxxxxxxxxxxxxxxxxxxxxxxxxxxxxxxxx
\begin{figure}[t!]
%\hspace{-2cm}\epsffile{fi.eps}
%\vspace{-6.2em}
\centerline{%
\epsfxsize=15cm%
\epsffile{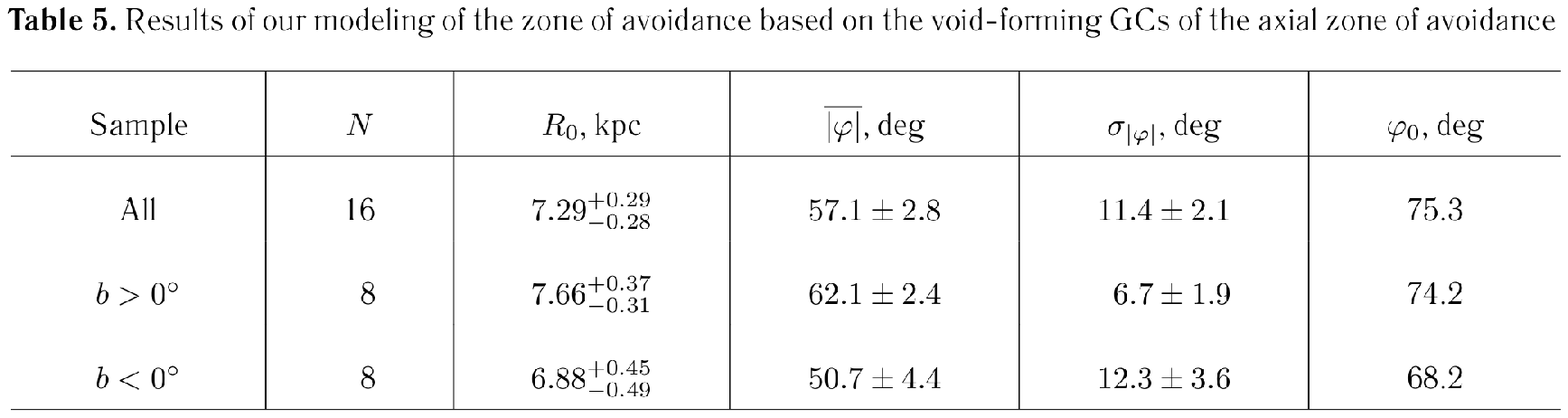}%
}%
%\vspace{-6em}
%\label{maxcone}
\end{figure}
%xxxxxxxxxxxxxxxxxxxxxxxxxxxxxxxxxxxxxxxxxxxxxxxxxxxxxxxxxxxxx

The solution was found for each of the three GC samples. The error of $R_0$ in all three
cases turned out to be quite low, despite the very small size of the samples. The
difference between the $R_0$ estimates based on the northern and southern samples is not
revealing due to its low confidence level ($1.4\sigma$) and the small size of the samples.
We took the estimate based on the complete sample of void-forming GCs as the final result
within this method: $R_0={7.29}^{+0.29}_{-0.28}$~kpc.  Figure 8a presents the profile of
the objective function for the parameter $R_0$ for the complete sample. The profile has a
local minimum at $R_0 = 9.34$~kpc, but it is shallow and is found at a $3.6\sigma$
(99.97\%) confidence level with respect to the global minimum. Thus, the confidence region
for the latter remains connected up to this level. The local minimum is attributable to
the southern GCs: it is also present and is deeper for the southern sample and is absent
for the northern one. 

%xxxxxxxxxxxxxxxxxxxxxxxxxxxxxxxxxxxxxxxxxxxxxxxxxxxxxxxxxxxxxx
\begin{figure}[t!]
%\hspace{-2cm}\epsffile{fi.eps}
%\vspace{-6.2em}
\centerline{%
\epsfxsize=15cm%
\epsffile{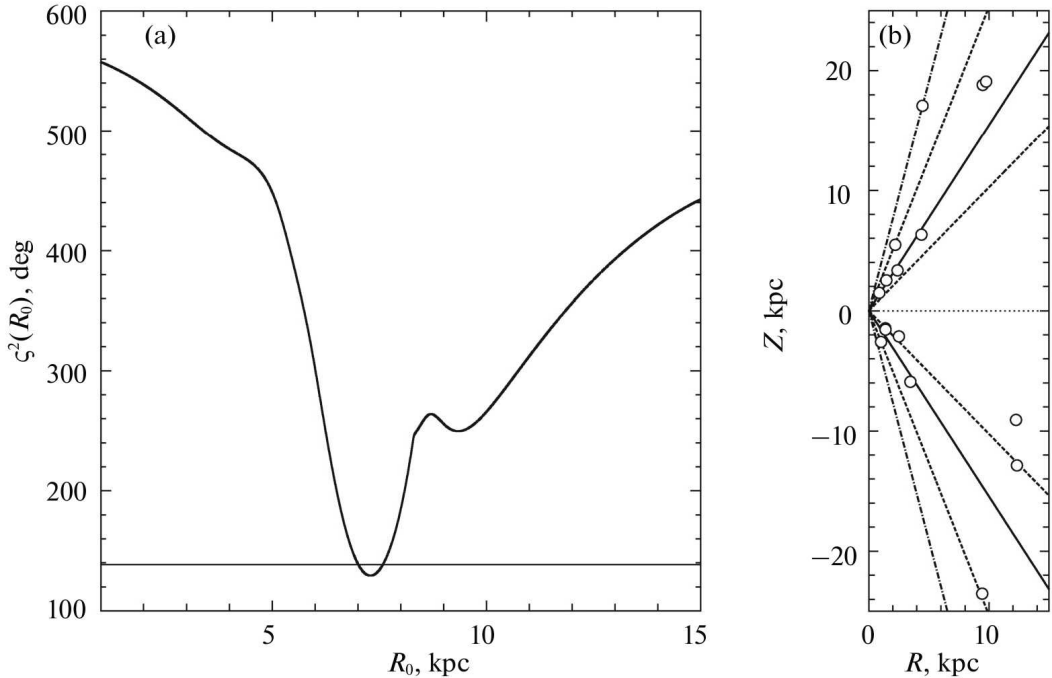}%
}%
%\vspace{-6em}
\caption{\rm
Solution of system (4) for the complete sample of void-formingGCs. (a) The profile of the
objective function ${\sigma}^2_{|\varphi|}$ for the parameter $R_0$\,. The horizontal line
marks the~$1\sigma$ confidence level. (b) The distribution of GCs from the sample in
coordinates $(R,Z)$ for $R_0$ = 7.29 kpc. The oblique solid, dashed, and dash--dotted
lines correspond to the ``model'' $\overline{|\varphi|}= 57\fdg1$,
$\overline{|\varphi|}\pm{\sigma}_{|\varphi|}$ , and the formal COA boundary
$\varphi_0=75\fdg3$, respectively.} %\label{maxcone}
\end{figure}
%xxxxxxxxxxxxxxxxxxxxxxxxxxxxxxxxxxxxxxxxxxxxxxxxxxxxxxxxxxxxx

Figure 8b for $R_0 = 7.29$~kpc shows the distribution of GCs from the complete sample in
coordinates $(R,Z)$ in comparison with the angular model characteristics. The criterion
for excluding the objects with excessive residuals (Nikiforov 2012) does not give grounds
to reject some of these clusters even in the most rigorous case ($L'=1$).

The estimates of the angular parameters based on the northern and southern samples differ
noticeably (Table 5). For example, the difference
$\overline{|\varphi|}_{\text{N}}-\overline{|\varphi|}_{\text{S}}=11\fdg4 \pm 5\fdg0$ is
marginally significant ($2.3\sigma$). Indeed, the distributions of northern and southern
GCs in Fig.~8b seem to be somewhat different. However, this does not necessarily imply
that the parameters of the northern and southern COA cavities are different, because those
southern GCs that are near the void boundaries far from the COA axis make a strong
contribution to these differences, while the COA is determined by the clusters near the
close boundaries (Figs. 7, 8b). In any case, these results are not an independent
confirmation of the north--south difference for the region of avoidance noted in Section
4, being obtained from the same sample of void-forming GCs. We will return to this
question after applying the second method, which does not require the selection of
clusters and is applicable to their complete sample.

Note that all values of $\overline{|\varphi|}$ in Table 5 exceed considerably and
significantly the mean $\mex\Phi=\pi/2-1\approx 32\fdg 7$ for a spherically symmetric GC
distribution without COA (see Appendix A1).
\medskip

%.............................................................
\subsection{Model Distributions of Galactocentric Latitudes}
\label{modelling_distributions}

Let us derive the differential distribution law of Galactocentric latitudes $f({\varphi})$
by assuming that the GCs are arranged spherically symmetrically relative to the Galactic
center but are completely absent in the COA with a
half-angle~$\alpha_0\equiv\pi/2-\varphi_0$ and an axis coincident with the Galactic axis.
The GC distribution function in Galactocentric Cartesian coordinates $x$, $y$, and $z$ is then
\begin{equation}
{\mathfrak f_1}(x,y,z)= \mathfrak f(R_\text{g}), \qquad
R_\text{g}=\sqrt{x^2+y^2+z^2}=\sqrt{(X-R_0)^2+Y^2+Z^2}.
\end{equation}  
Here, $R_\text{g}$ is the
Galactocentric distance; $X$, $Y$, and $Z$ are the heliocentric Cartesian coordinates.
Let us introduce the spherical coordinates $R_\text{g}$, $\varphi$, and $\theta$: 
\begin{equation}
x=R_\text{g}\cos{\varphi}\cos{\theta}, \qquad%\nonumber 
y=R_\text{g}\cos{\varphi}\sin{\theta}, \qquad%\nonumber 
z=R_\text{g}\sin{\varphi}. %\nonumber
\end{equation}  
For the distribution function in coordinates $R_\text{g}$, 
$\varphi$, and $\theta$ we then have 
\begin{equation}
{\mathfrak f_2}(R_\text{g},\varphi,\theta)\,d{R_\text{g}}\,d{\varphi}\,d{\theta}=
{\mathfrak f_1}(x,y,z)\,d{x}\,d{y}\,d{z}=
{\mathfrak f_1}(R_\text{g},\varphi,\theta)J\,d{R_\text{g}}\,d{\varphi}\,d{\theta},
\end{equation}  
where $J=R^{2}_\text{g}\,\cos{\varphi}$ is the Jacobian (see, e.g., Agekyan 1974). Hence we
obtain
\begin{equation}
\label{fk_f2}
{\mathfrak f_2}(R_\text{g},\varphi,\theta)={\mathfrak f}(R_\text{g})\,{R^2_\text{g}}\,\cos{\varphi} . 
\end{equation}  
Integrating (10) over $R_\text{g}$ and $\theta$ gives an
expression for $f(\varphi)$ to within the normalization constant $c$:
\begin{equation}
\label{fr}
f(\varphi)=2\pi c I \cos{\varphi} ,\qquad
I=\int\limits^{a}_{0}{\mathfrak f}(R_\text{g})\,{R^2_\text{g}}\,dR_{\text{g}}\,.
\end{equation}
The normalization condition
\begin{equation}
1=\int\limits^{\pi/2}_{-\pi/2}{f(\varphi)\,d\varphi}=
2\pi c I\int\limits^{\varphi_0}_{-\varphi_0}{\cos{\varphi}\,d\varphi}=
4\pi c I \sin{\varphi_0}
\end{equation}  
defines the constant
\begin{equation}
\label{c}
    c=\frac{1}{4\pi I \sin{\varphi_0}}\,. 
\end{equation}  
Using (11) and (13), we obtain the sought-for model
distribution of angles $\varphi$ for a spherically symmetric
GC distribution with an axial COA:
\begin{equation}
\label{1model}
f({\varphi})=\left\{
\begin{array}{ll}
\frac{\displaystyle\cos{\varphi}}{\displaystyle 2\sin{\varphi}_0}\,,\,&|{\varphi}|\leq{\varphi}_0\,,\\
0,        &  |{\varphi}|>{\varphi}_0\,.
\end{array}
\right. 
\end{equation}
For a strictly spherically symmetric GC distribution
(without COA, $\varphi_0=\pi/2$) the differential law (14)
takes the form
\begin{equation}
\label{1model_noCoA}
f({\varphi})=\frac{\cos{\varphi}}{2}\,. %\nonumber
\end{equation}

However, directly applying the simple (single-component) model (14) leads to a number of
difficulties. For example, even if this model is ideal, the fall of at least one GC into
the ``forbidden zone''~$|\varphi|>\varphi_0$ during the model optimization at trial values
of the model parameters can lead to a formally infinite value of the objective function.
However, even if the parameters of model (14) are correct and if the latter is completely
adequate, a GC can formally end up in the zone~$|\varphi|>\varphi_0$ due to the random
error in the heliocentric distance~$r$, which is quite probable for GCs with small
$|\varphi|$ in the region $(X,Y)\approx (R_0,0)$. Note that this effect is insignificant
for clusters with large $|\varphi|$ by which the COA is actually identified (Section 4). In
addition, the possibility that some GC is physically in the COA must not be ruled out,
because it fell there comparatively recently and the COA-cleaning factors have not yet
affected it. Finally, the COA may not exist in reality for the central GCs. This is
suggested by the above-mentioned results of numerical experiments by Sasaki and Ishizawa
(1978) and by the fact that the axial zone of avoidance in the layers at small $|Z|$
cannot be identified by the observed GC distribution (see Section 4).

All these factors can be taken into account to a
first approximation if we envisage a component of
the GC system without COA in the model. In what
follows, we will consider a two-component model of
the Galactocentric latitude distribution function:
\begin{equation}
\label{f2}
f_2({\varphi})=f_{\text{b}}({\varphi})+f_{\text{c}}({\varphi}),\quad
f_{\text{c}}({\varphi})=\left\{
\begin{array}{ll}
\frac{\displaystyle{C\cos{\varphi}}}{\displaystyle2\sin{\varphi}_0},\,& |{\varphi}|\leq{\varphi}_0\,,\\
0,        & |{\varphi}|>{\varphi}_0\,,
\end{array}
\right. \enskip
f_{\text{b}}({\varphi})=\frac{(1-C)\cos{\varphi}}{2}\,.  %\nonumber
\end{equation}
Here, $f_\text{b}$ is the bulge component without COA (``b''
stands for bulge), $f_\text{c}$ is the component with COA
(``c'' here stands for cone), and $C$ is the GC fraction
accounted for by the latter component ($1>C\geq0$).
The model (16) is also convenient from a computational
point of view, because at $C < 1$ it does not
rule out {\em completely\/} the fall of clusters into the region
$|\varphi|>\varphi_0$\,.
\medskip

%.............................................................
\subsection{Modeling without A priori Information about the
Cone of Avoidance}
\label{modelling_results}

Comparison of model (16) with the data on GCs constitutes the second modeling method.
Optimization of the vector of model parameters $\mathbf a$ [in the general case, $\mathbf
a=(R_0, {\varphi}_0, C)$] allows the solution to be obtained without any additional
information about the COA for an arbitrary GC sample produced without using any a priori
(initial) values of the COA parameters, including those for the complete GC sample.

We sought for the values of the parameters that minimized the statistics
\begin{equation}
{\chi}^2(\mathbf a)={\sum\limits^{n_\text{b}}_{i=1}{\frac{({\nu}_i-N{p}_i)^2}{N{p}_i}}}\,. %\nonumber
\end{equation}
Here, $N$ is the total number of objects in the sample;
$n_\text{b}$ is the number of bins into which the interval
$[-\pi/2,\pi/2]$ of possible $\varphi$ was divided; ${\nu}_i$ is the
number of objects that fell into the $i$th bin; ${p}_i$ is the
probability that an object falls into the $i$th bin,
\begin{equation}
\label{binning}
p_i=\int\limits^{\varphi_i+\Delta\varphi_i}_{\varphi_i}{f_2(\varphi(R_0); \varphi_0,C)\,d\varphi},
\quad i=1,\:\ldots,\:n_\text{b}\,,
\end{equation}
where $f_2(\varphi)$ is the model distribution (16), and $\Delta\varphi_i$
is the width of the $i$th bin. After our trial calculations
we chose a constant~$\Delta\varphi=10\deg$ for all bins that, on
average, corresponds to the general recommendation
$\nu_i\ge 5$ (Sveshnikov 2008).

The confidence intervals for the parameters $R_0$ and
$\varphi_0$ were found by a method similar to that applied in
Subsection 5.1. In the case of the $\chi^2$ statistics, the
boundaries of the confidence interval of the parameter
$a_j$ for a $k\sigma$ confidence level are given by the equation
\begin{equation} 
    \chi^2_{\text{m}}(a_j)=\chi^2_k\equiv\chi^2_0 + k^2,
\end{equation} 
where
\begin{equation} 
%\label{}
    \chi^2_0\equiv \min\chi^2 (\mathbf a),
    \qquad
    \chi^2_{\text{m}}(a_j)\equiv \min_{a_j\,=\:\text{const}} \chi^2 (\mathbf a)
\end{equation} 
(Press et al.\ 1997). Figure 9 presents the $\chi^2_{\text{m}}(R_0)$ and
$\chi^2_{\text{m}}(\varphi_0)$ profiles for the complete GC sample.

%xxxxxxxxxxxxxxxxxxxxxxxxxxxxxxxxxxxxxxxxxxxxxxxxxxxxxxxxxxxxxx
\begin{figure}[t!]
%\hspace{-2cm}\epsffile{fi.eps}
%\vspace{-6.2em}
\centerline{%
\epsfxsize=15cm%
\epsffile{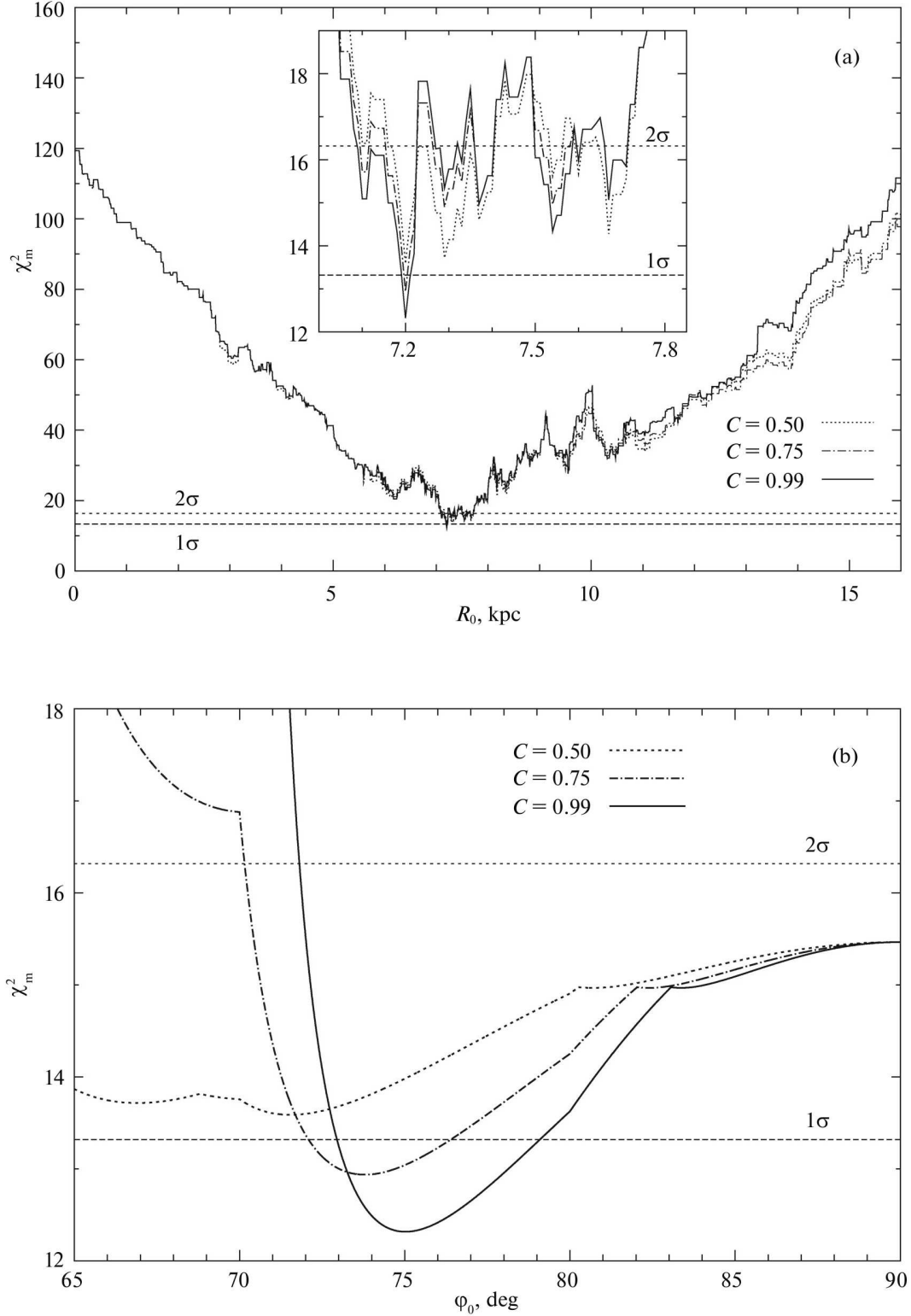}%
}%
%\vspace{-6em}
\caption{\rm
Profiles of the objective function $\chi^2$ for the parameters $R_0$ (a) and~$\varphi_0$
(b) at fixed values of the fraction of the component of the GC system with COA $C=0.50$,
$0.75$, $0.99$ for the complete sample of clusters. The horizontal lines mark the~$1\sigma$
and $2\sigma$ confidence levels for the solution at $C = 0.99$.}
\end{figure}
%xxxxxxxxxxxxxxxxxxxxxxxxxxxxxxxxxxxxxxxxxxxxxxxxxxxxxxxxxxxxx

Our attempts to solve the complete problem of optimizing the parameters of model~(16) for
different GC subsamples showed that $C\to1$; the objective function acquires a distinct
minimum (i.e., an unambiguous solution exists) only when the component with COA dominates
($C > 0.5$) (Fig.~9). These results argue for the existence of COA in the GC system. Below
we adopt $C = 0.99$: as $C$ increases further, the estimates of the parameters barely
change, but the confidence level at which the confidence region loses its connectivity
slightly lowers.

Owing to the binning of the domain of definition of $\varphi$ for calculating the
probabilities~(18) and numbers~$\nu_i$ and the presence of a sharp truncation at
$|{\varphi}|={\varphi}_0$ in the component~$f_{\text{c}}({\varphi})$ of model~(16), the
objective function~$\chi^2(R_0,\varphi_0)$ turns out to be nonsmooth. For example,
the~$\chi^2_{\text{m}}(R_0)$ profile (Fig.~9a) is, strictly speaking, a piecewise constant
function, although the segments with a constant~$\chi^2_{\text{m}}$ value near the global
minimum are quite small (usually have a length~$\Delta R_0\la 0.01$~kpc). In addition, the
function~$\chi^2_{\text{m}}(R_0)$ experiences shallow but sharp oscillations. However, the
solution of the problem is given by the narrow (the formal~$1\sigma$ errors are
$\sigma_{R_0}=0.01$--$0.04$~kpc) and fairly deep (unique up to the $1.4\sigma$ confidence level)
minimum (Fig.~9a, Table 6).

%xxxxxxxxxxxxxxxxxxxxxxxxxxxxxxxxxxxxxxxxxxxxxxxxxxxxxxxxxxxxxx
\begin{figure}[t!]
%\hspace{-2cm}\epsffile{fi.eps}
%\vspace{-6.2em}
\centerline{%
\epsfxsize=15cm%
\epsffile{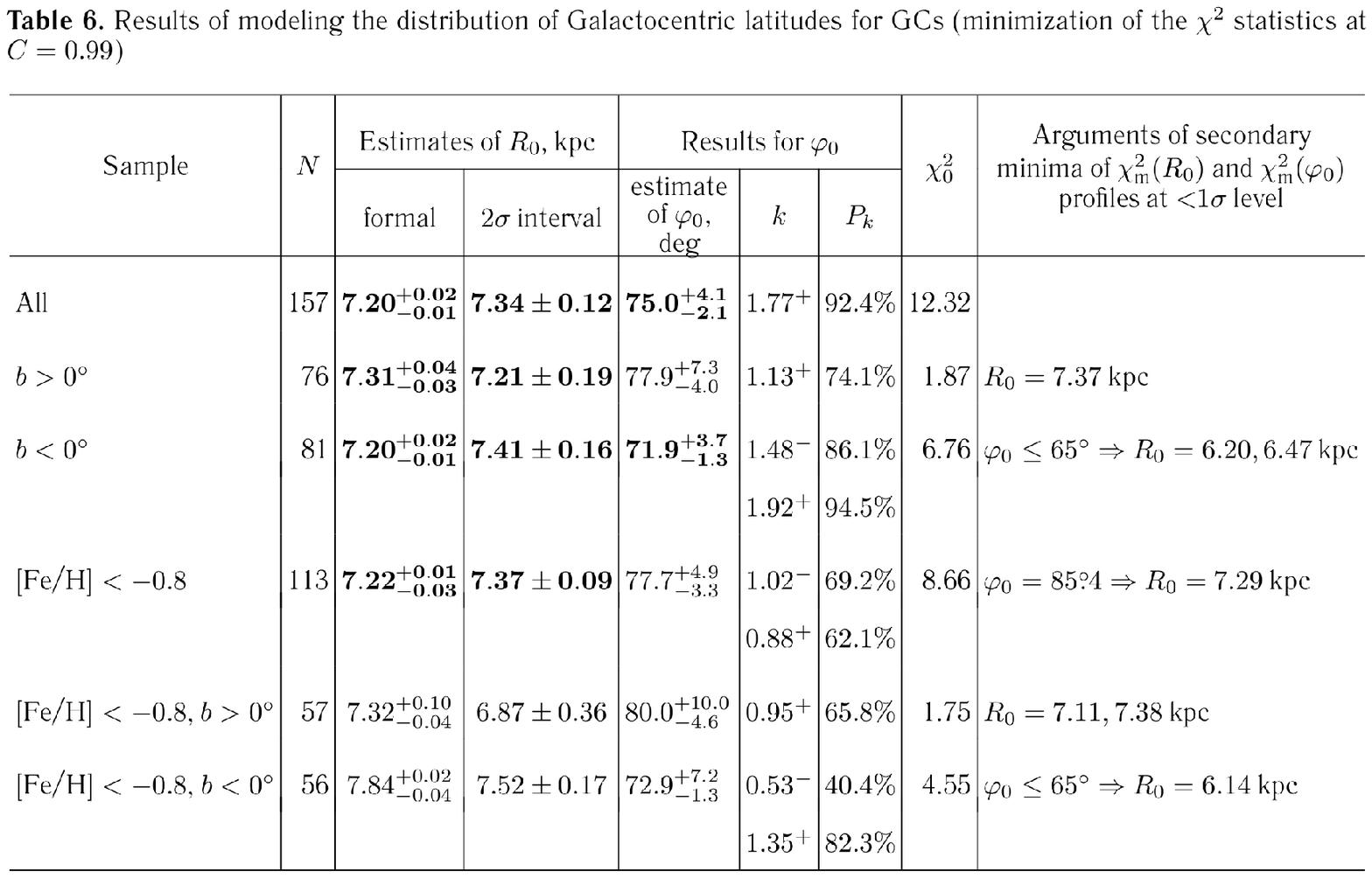}%
}%
%\vspace{-6em}
%\label{maxcone}
\end{figure}
%xxxxxxxxxxxxxxxxxxxxxxxxxxxxxxxxxxxxxxxxxxxxxxxxxxxxxxxxxxxxx

The depth of the minimum in the~$\chi^2_{\text{m}}(\varphi_0)$ profile is limited by the
difference between the value of this function at the right boundary of the domain of
definition of $\chi^2_{\text{m}}(\pi/2)$, which does not depend on $C$, and the $\chi^2_0$
value as a function of $C$ (Fig.~9b). The significance of the minimum turns out to be only
close to the marginal one: for example, for the complete GC sample it is $1.8\sigma$
(92.4\%) at $C = 0.99$ and grows weakly as $C$ increases further. The
parameter~$\varphi_0$ is limited from below much more strongly than from above (Fig.~9b).
Note that when the COA is abandoned in the model or when the contribution of the component
with COA is simply reduced greatly, the determination of the parameter~$R_0$ becomes
ambiguous (see the $\chi^2_{\text{m}}(R_0)$ profile at $C = 0.5$ in Fig.~9a).

The method was applied to the complete GC sample, to the northern ($b>0^\circ$) and
southern ($b<0^\circ$) GCs separately, to the GCs with $[\text{Fe/H}]<-0.8$, and to the
samples with combinations of these constraints. The constraint $[\text{Fe/H}]<-0.8$ stems
from the fact that the higher-metallicity GCs form an oblate subsystem and can shift the
results for a spherically symmetric model. The results obtained for these samples at $C =
0.99$ are summarized in Table 6; the formal errors of the parameters for the~$1\sigma$
level are given. Figure 10 presents the distributions of Galactocentric latitudes for the
complete sample and the GC sample with $[\text{Fe/H}]<-0.8$ in comparison with the
models~(16) constructed for them and with the model without COA (15) at the optimal
parameters $R_0$ and~$\varphi_0$ (providing a minimum of the function $\chi^2$). Figure 10
illustrates good agreement of the observed $\varphi$ distributions with the model ones.
Applying Pearson's test leads to the same conclusion: for all of the samples considered at
the optimal parameters $\chi_0^2<N_{\text{free}}$\,, where $N_{\text{free}}=16$ for the GC
samples without any constraints on $b$, $N_{\text{free}}=7$ for the samples of northern or
southern GCs (see, e.g., Press et al.\ 1997).

%xxxxxxxxxxxxxxxxxxxxxxxxxxxxxxxxxxxxxxxxxxxxxxxxxxxxxxxxxxxxxx
\begin{figure}[t!]
%\hspace{-2cm}\epsffile{fi.eps}
%\vspace{-6.2em}
\centerline{%
\epsfxsize=0.67\textwidth
\epsffile{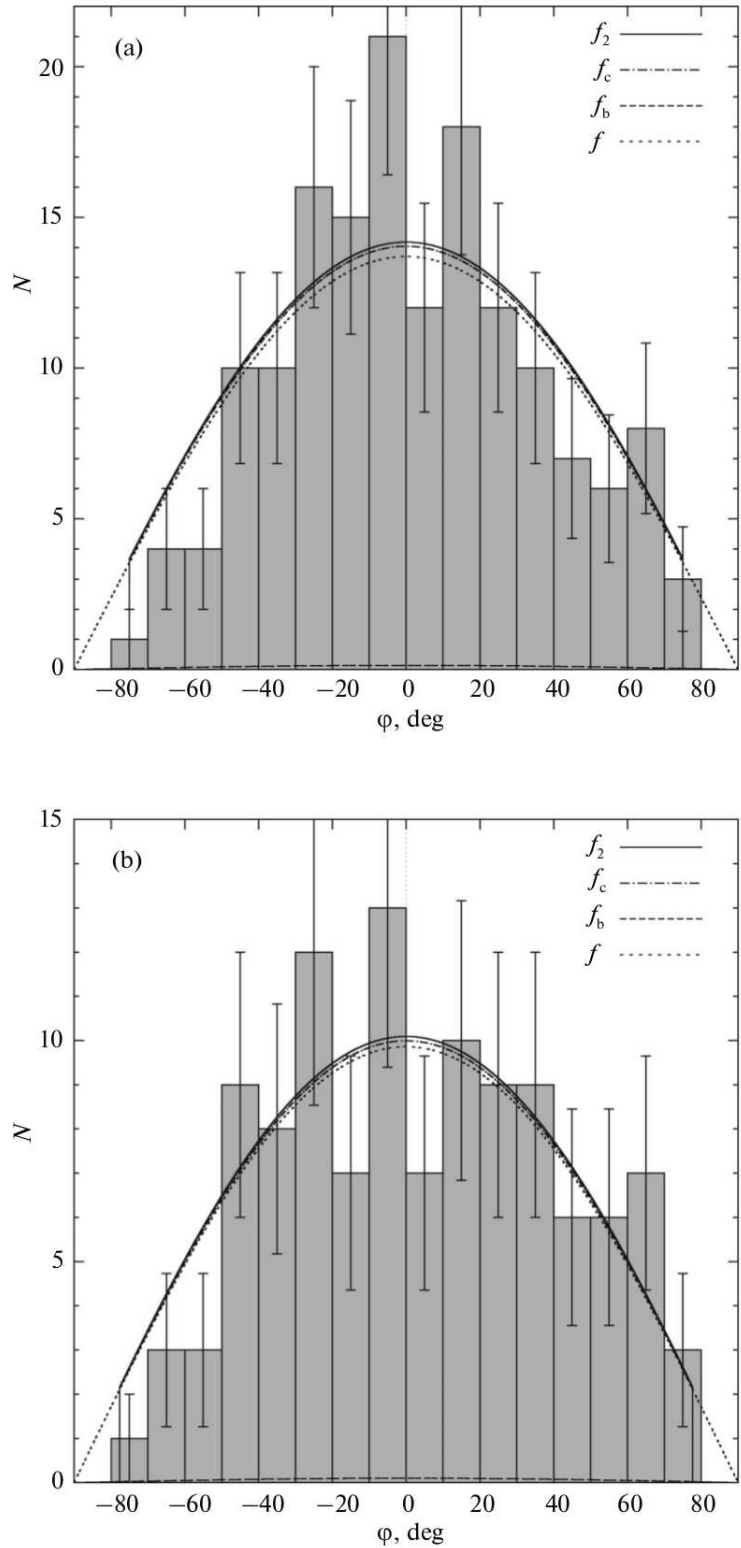}%
%\epsffile{fig10_100_50.eps}%
%\epsffile{fig10.eps}%
}%
%\vspace{-6em}
\caption{\rm
Distribution of GCs in Galactocentric latitudes in comparison with the
model~$f_2(\varphi)$ that includes the component~$f_\text{c}$ with COA and the component
$f_\text{b}$ without COA and with the model~$f(\varphi)$ in the absence of COA. $R_0$
and~$\varphi_0$ are the values minimizing $\chi^2$ at $C = 0.99$. (a)~The~complete sample
($N = 157$), $R_0 = 7.20$~kpc, $\varphi_0=75\fdg0$. (b)~The~sample with
$[\text{Fe/H}]<-0.8$ ($N = 113$), $R_0 = 7.22$~kpc, $\varphi_0=77\fdg7$.}
\end{figure}
%xxxxxxxxxxxxxxxxxxxxxxxxxxxxxxxxxxxxxxxxxxxxxxxxxxxxxxxxxxxxx

Apart from the formal estimates of $R_0{\color{black}(\chi^2_0)}$ corresponding to the
global minimum of~$\chi^2$, we also found the {\em interval\/} estimates of this parameter
whose derivation we consider as a method of smoothing the oscillations of the
function~$\chi^2_{\text{m}}(R_0)$. Here, we proceed from the following reasoning. The
dependence $\chi^2_{\text{m}}(R_0)$ has several local minima below the~$\chi^2_2$~value
($2\sigma$ level). Random factors, for example, errors in the distances, can determine
precisely which of them will be deepest. A characteristic feature of the function
$\chi^2_{\text{m}}(R_0)$ is its oscillation relative to the ${\approx}2\sigma$ level in
the region of the global minimum in a comparatively narrow interval of $R_0$ outside which
$\chi^2_{\text{m}}$ increases sharply (Fig.~9a). This gives grounds to rely on the
confidence interval for the $2\sigma$ level when estimating $R_0$\,. First we found the
set of local minima of $\chi^2_{\text{m}}(R_0)$ smaller than $\chi^2_2$ for which the
values of~$\varphi_0$ corresponding to them fell within the formal~$1\sigma$ interval
for~$\varphi_0$ determined from the dependence~$\chi^2_{\text{m}}(\varphi_0)$. Then, we
determined the intervals of $R_0$ in which the curves of these local minima did not exceed
$\chi^2_2$\,. The largest and smallest boundaries of these intervals were taken as the
boundaries of the $2\sigma$ interval for estimating $R_0$\,. We took the middle of the
interval found in this way as the interval estimate of $R_0{\color{black}(2\sigma)}$ and a
quarter of its length as the~$1\sigma$ error of this estimate. The interval estimates of
$R_0$ are also given in Table 6.

Next, among the estimates of the parameters obtained we revealed obviously unreliable
ones. These included the $R_0$ and~$\varphi_0$ estimates based on the samples for which
the function $\chi^2_{\text{m}}(\varphi_0)$ did not reach the $1\sigma$~level at least on
one side of the point estimate at~$\varphi_0\ge 65^\circ$ (the subsamples of northern and
southern low-metallicity GCs). [There were no such cases for the~$\chi^2_{\text{m}}(R_0)$
profiles.] The~$\varphi_0$ estimates with a low significance of the minimum
of~$\chi^2_{\text{m}}(\varphi_0)$ (with a confidence probability $P_k < 85\%$) were also
considered as unreliable ones. The values of $P_k$ corresponding to a $k\sigma$ level,
$k=\sqrt{\chi^2_k-\chi^2_0}$\,, were determined from the values of the
function~$\chi^2_{\text{m}}(\varphi_0)=\chi^2_k$ at the boundaries of the
interval~($\varphi_0=65^\circ, 90^\circ$) under consideration or at the points of its
local maxima rightward and leftward of the point estimate. The values of $k$ and $P_k$ are
given in Table 6, where ``$+$'' and ``$-$'' at $k$ denote the points of
$\chi^2_{\text{m}}(\varphi_0)$ rightward and leftward of the point estimate
of~$\varphi_0$\,, respectively.

The remaining (more reliable) estimates of the parameters are highlighted in Table~6 in
boldface. Based on them, we can draw the following conclusions with regard to the
parameter $R_0$\,. The formal and $2\sigma$-interval estimates are close ($|\Delta R_0|\le
0.21$~kpc), without any clear shift of some relative to the others. The formal uncertainty
in the estimates turned out to be low (within 0.2 kpc), but, as our numerical experiments
showed, it was underestimated, especially for $R_0(\chi_0^2)$. The $R_0$ estimates based
on the northern and southern GCs and on the complete sample differ by no more than
0.2~kpc. Taking into account the more realistic errors of the interval estimates, this
difference is insignificant. Besides, the sign of the north--south difference in the~$R_0$
estimates is different for the formal and interval estimates. In any case, there is no
evidence for $R_0$ estimated from the southern GCs being underestimated in comparison with
those from the northern ones. Obviously, the inverse result formally obtained in
Subsection 5.1 based on small samples of void-forming clusters is random in nature. The
exclusion of metal-rich GCs barely changes the $R_0$ estimates.

The $R_0$ estimates highlighted in Table 6 were averaged with weights inversely
proportional to the squares of the lengths of the confidence intervals. Since these
estimates are not independent, as the error of the mean we took the square root of the
weighted mean value of the squares of the errors in the estimates being averaged. By the
error here we mean the positive or negative parts of the confidence interval for
the~$1\sigma$ level if they are different. We found the mean of four formal estimates,
$\overline{R_0{\color{black}(\chi^2_0)}}=7.21\pm0.01$~kpc, and the mean of the same number
of interval estimates, $\overline{R_0{\color{black}(2\sigma)}}=7.35\pm0.10$~kpc. The
difference in $\overline{R_0}$ for the two variants of estimates is insignificant.

This approach was tested by the Monte Carlo method. For each initial model we generated
1000 pseudo-random catalogs of GCs distributed spherically symmetrically with a radial
density law~${\mathfrak f}(R_\text{g})\propto R_\text{g}^{-2.5}$ (Rastorguev et al.\ 1994)
in the absence of COA ($\varphi_0=90^\circ$) and in its presence. The actual region of
avoidance apparently goes beyond the COA, being nonaxisymmetric with a high probability
(see Subsection 6.4), which manifests itself in a strong deficit of GCs seen within the
COA in projection onto the  $XZ$ plane compared to its projection onto the $YZ$  plane
(Fig.~11). Therefore, we also considered a limiting case of this deficit as a model: the
absence of GCs in the ``trough of avoidance'' $|\beta|>\beta_0$, where $\beta$ is the
Galactocentric angular elevation of GCs above the Galactic plane in projection onto the
$XZ$ plane. In the models we adopted $R_0 = 7.2$~kpc (the mean estimate in recent papers
on GCs; see Table 8); $N$ and~$\varphi_0$ were taken to correspond to the complete sample
and the sample with $[\text{Fe/H}]<-0.8$ (Table 6). We took $\beta_0$ to be equal
to~$\varphi_0$ or $76^\circ$ (it corresponds to the largest value of the mentioned deficit; see
Subsection 6.4). The results are presented in Table 7.

%xxxxxxxxxxxxxxxxxxxxxxxxxxxxxxxxxxxxxxxxxxxxxxxxxxxxxxxxxxxxxx
\begin{figure}[t!]
%\hspace{-2cm}\epsffile{fi.eps}
%\vspace{-6.2em}
\centerline{%
\epsfxsize=15cm%
\epsffile{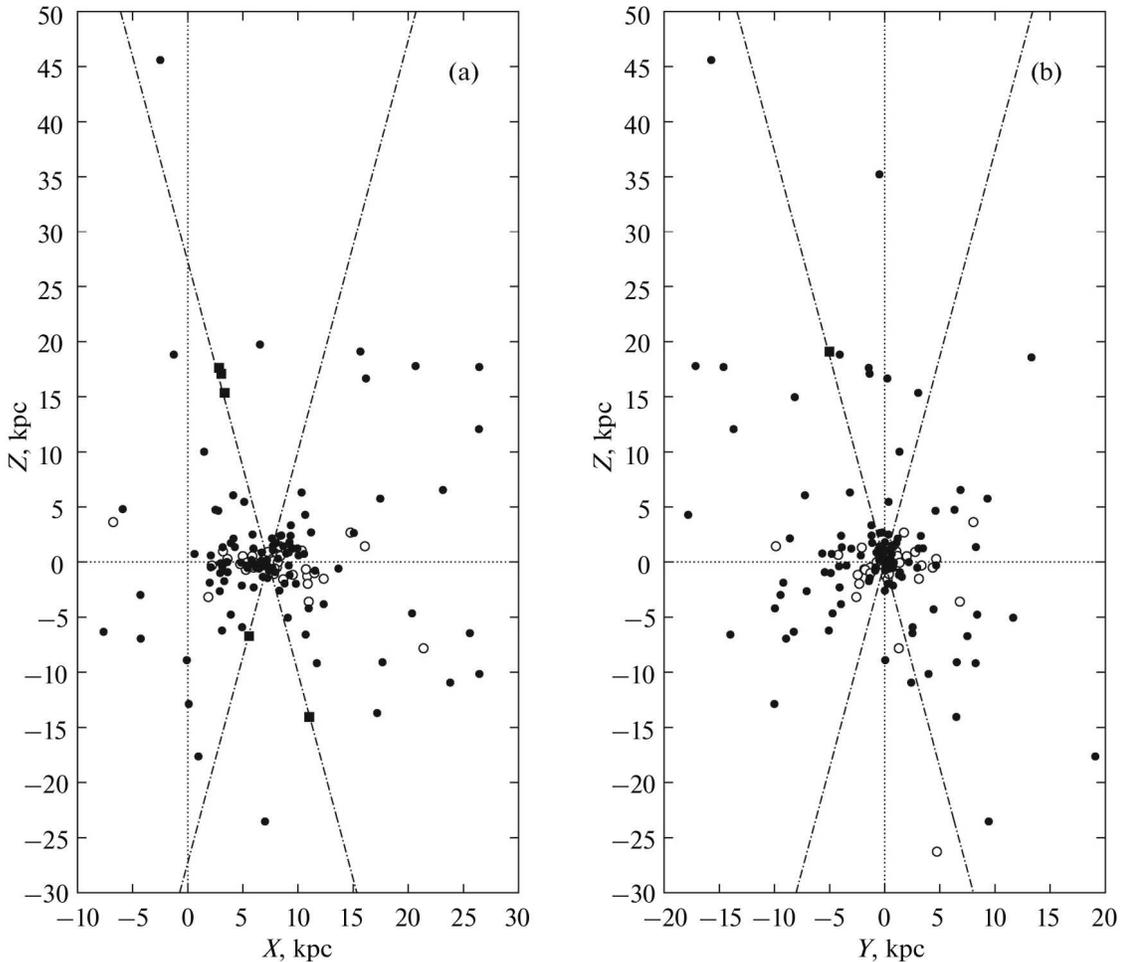}%
}%
%\vspace{-6em}
\caption{\rm
Distribution of GCs and the contour (dash--dotted lines) of the COA ($\alpha_0=15\fdg 0$,
$R_0=7.{\color{black}3}$~kpc) in projection onto the  $XZ$ (a) and $YZ$ (b) planes. The
open circles are the clusters with $\text{[Fe/H]}>-0.8$; the filled symbols are the
remaining clusters. The squares mark the clusters that are projected onto the COA cavity
at $\alpha_0=15\fdg 0$ but are outside the COA at $\alpha_0={\color{black}14\fdg 0}$. For
each of the two projections there is one cluster that is projected onto the COA with
$\alpha_0=15\fdg 0$ but is outside the figure.} %\label{maxcone}
\end{figure}
%xxxxxxxxxxxxxxxxxxxxxxxxxxxxxxxxxxxxxxxxxxxxxxxxxxxxxxxxxxxxx

Our numerical experiments showed both estimates of $R_0(\chi^2_0)$ and
$R_0{\color{black}(2\sigma)}$ to be unbiased. The variance of the estimates strongly
depends on the assumptions about the region of avoidance. The expectation that the
presence of such a region made the $R_0$ estimates more effective was confirmed. The
uncertainty in the $R_0{\color{black}(2\sigma)}$ estimates in the presence of a region of
avoidance is higher than the uncertainty in the $R_0(\chi^2_0)$ estimates approximately by
30\%. The formal errors in $R_0(\chi^2_0)$ turned out to be clearly inadequate
(underestimated approximately by an order of magnitude), obviously because the objective
function is nonsmooth. The assumption that the formal errors in the interval estimates
were more realistic was confirmed, but they were also underestimated.

Averaging the estimates based on real data $\overline{R_0(\chi^2_0)}$ and
$\overline{R_0(2\sigma)}$ with weights of $(1.3)^2$ and $1$, respectively, gives
$\overline{R_0}=7.26$~kpc. Given that the actual region of avoidance is closer to the
``trough'' model (Section 6.4) and that $N = 113$ corresponds to the sample with
$[\text{Fe/H}]<-0.8$ more consistent with the assumption of spherical symmetry, based on
our numerical experiments we take the uncertainty in this estimate to be $\pm0.5$~kpc.
Finally, we obtain an estimate of $\overline{R_0}=7.3\pm0.5$~kpc by this method.

Table 7 shows that the parameter~$\varphi_0$ in our numerical
experiments is reconstructed within the limits of
errors whose level is close to the formal ones (Table 6).

%xxxxxxxxxxxxxxxxxxxxxxxxxxxxxxxxxxxxxxxxxxxxxxxxxxxxxxxxxxxxxx
\begin{figure}[t!]
%\hspace{-2cm}\epsffile{fi.eps}
%\vspace{-6.2em}
\centerline{%
\epsfxsize=9cm%
\epsffile{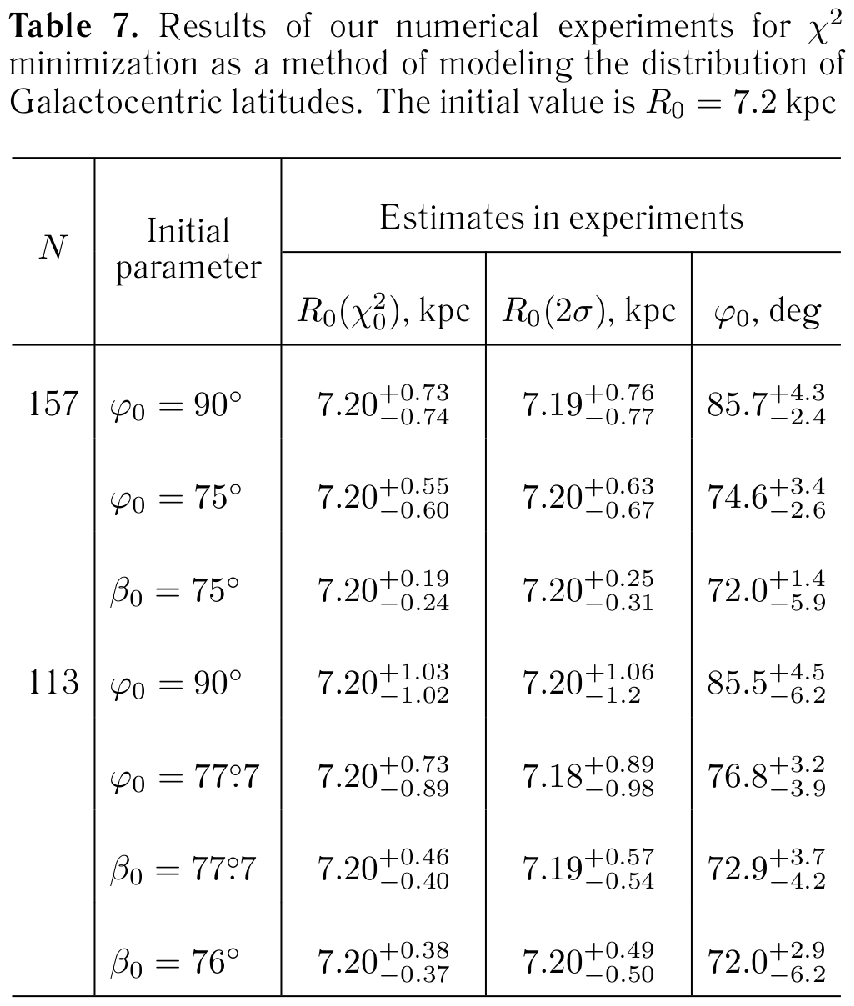}%
}%
%\vspace{-6em}
%\label{maxcone}
\end{figure}
%xxxxxxxxxxxxxxxxxxxxxxxxxxxxxxxxxxxxxxxxxxxxxxxxxxxxxxxxxxxxx

Although all of the approaches implemented in this paper lead to a positive difference
of~$\varphi_0$ from the northern and southern GCs, $\Delta
\varphi_0\equiv\varphi_{0,\text{N}}-\varphi_{0,\text{S}}=+(6\deg$--$8\deg)$ (Tables 1--6),
this difference is found to be insignificant when the distribution of Galactocentric
latitudes is modeled directly. The most reliable results (for the samples with $b>0^\circ$ 
and $b<0^\circ$) give $\Delta \varphi_0=+6\fdg0\pm 4\fdg2$ ($1.4\sigma$). (Note that for
the northern GCs the significance of the minimum of~$\chi^2_{\text{m}}(\varphi_0)$ is low
only in the sense of a constraint on~$\varphi_0$ from above, but not from below.) Thus, we
have no reason to believe that the northern and southern COA cavities differ significantly
in angular size. In Fig.~10 we can see some deficit of southern GCs at~$\varphi\la-50\deg$.
However, the Kolmogorov--Smirnov test shows that the distributions of southern and
northern GCs in $\varphi$ differ insignificantly: the null hypothesis is rejected only at
the 68 and 77\% levels for the complete sample and the sample of metal-poor GCs,
respectively.

The estimate of~$\varphi_0=75\fdg0^{+4\fdg1}_{-2\fdg1}$ based on all GCs was taken as the
final one as having the greatest significance~$P_k$ of the minimum of
$\chi^2_{\text{m}}(\varphi_0)$. Averaging the three estimates of~$\varphi_0$ with the
greatest $P_k$ (based on all GCs and on the samples with $b>0\deg$  and $b<0\deg$) using
the same procedure as that for $R_0$ gives a smaller value,
$\overline{\varphi_0}=73\fdg6^{+4\fdg1}_{-1\fdg6}$, because the estimate based on the
southern GCs ($\varphi_0={71\fdg9}^{+3\fdg7}_{-1\fdg3}$) is formally more accurate.
However, the latter is not revealing, because a random shift of the~$\varphi_0$ estimate
to {\em smaller\/} values must lead to a larger formal conditionality: from general
considerations a wider COA is identified with greater confidence than a narrower one at
the same GC sample size (see also Table 6).
\medskip

%***************************************************************
\section{DISCUSSION}

The approaches considered in this paper (Subsections
4.2, 4.3, 5.1, 5.3) yield, on the whole, similar
results. Let us discuss what conclusions they allow
to draw and what further prospects for this method
are.
\medskip

%.............................................................
\subsection{The Existence of an Axial Zone of Avoidance}

The existence of such a zone in the Galactic GC system is revealed when analyzing the
distribution of locally maximal cylindrical voids (Section 4) and is confirmed by the
results of modeling the distribution of Galactocentric latitudes (Subsection 5.3). Within
the latter approach the $\chi^2$ minimization requires the dominance of the component with
COA and the solution of the problem turns out to be unambiguous only under the condition
of such dominance. The distinct minima of $\chi^2_{\text{m}}(\varphi_0)$ for the most
reliable solutions also confirm the existence of COA.

The probability that for a spherically symmetric GC distribution, i.e., for the
distribution law (15), none of $N$ GCs falls into a double COA with a halfangle $\alpha_0$
is $P_N(0)=\cos^N \alpha_0$, because $2\int_0^{\pi/2-\alpha_0}\frac{\cos \varphi}{2}\,
d\varphi=\cos \alpha_0$\,. Then, $P_{157}(0)=0.43\%$  and $P_{118}(0)=1.7\%$ (the sample
without GCs with $\text{[Fe/H]}>-0.8$ having an oblate distribution along the $Z$ axis).
However, it is not sufficient to rely only on the statistics of the $\varphi$ distribution
at some fixed $R_0$ in this question, because Pearson's test applied to such a
distribution does not reject the alternative model, the absence of COA, either: the
probabilities to obtain the observed or larger deviations from such a model are 51 and
87\% for the distributions on the upper and lower panels of Fig.~10, respectively,
although the $\chi^2$ statistics in the absence of COA is nevertheless poorer (16.23 and
10.69, respectively) than that in its presence (12.32 and 8.66). At the same time, the
value of $R_0$ obtained through general optimization is of crucial importance.

Using the results of our numerical experiments (Subsection 5.3), we can test the null
hypothesis by determining, in the absence of a region of avoidance, the probability to
obtain the same COA as that from real data or wider and, at the same time, a value of
$R_0$ as close as that from real data or closer to the mean $\langle
R_0\rangle_{\text{GC}}$ from papers based on an analysis of the spatial GC distribution
(see Table 8). For $\langle R_0\rangle_{\text{GC}}=7.18$~kpc (from 1989--2014 papers)
these probabilities at $N = 113$ are 1.1 and 2.1\% for $R_0(\chi^2_0)$ and
$R_0{\color{black}(2\sigma)}$, respectively; for $\langle R_0\rangle_{\text{GC}}=7.32$~kpc
(from 1975--2014 papers) they are 1.6 and 1.0\%. At $N = 157$ these probabilities are even
lower: ${<}0.0$, 0.4, 0.5, and 0.1\%, respectively. Thus, the hypothesis about the absence
of a region of avoidance is rejected at least at the 98\% level. Note that in the presence
of a cone or trough of avoidance the hypotheses about obtaining similar results by chance
are not rejected (the probabilities even for $N = 113$ turn out to be in the intervals
5.4--16 and 18--41\%, respectively).

%xxxxxxxxxxxxxxxxxxxxxxxxxxxxxxxxxxxxxxxxxxxxxxxxxxxxxxxxxxxxxx
\begin{figure}[t!]
%\hspace{-2cm}\epsffile{fi.eps}
%\vspace{-6.2em}
\centerline{%
\epsfxsize=16cm%
\epsffile{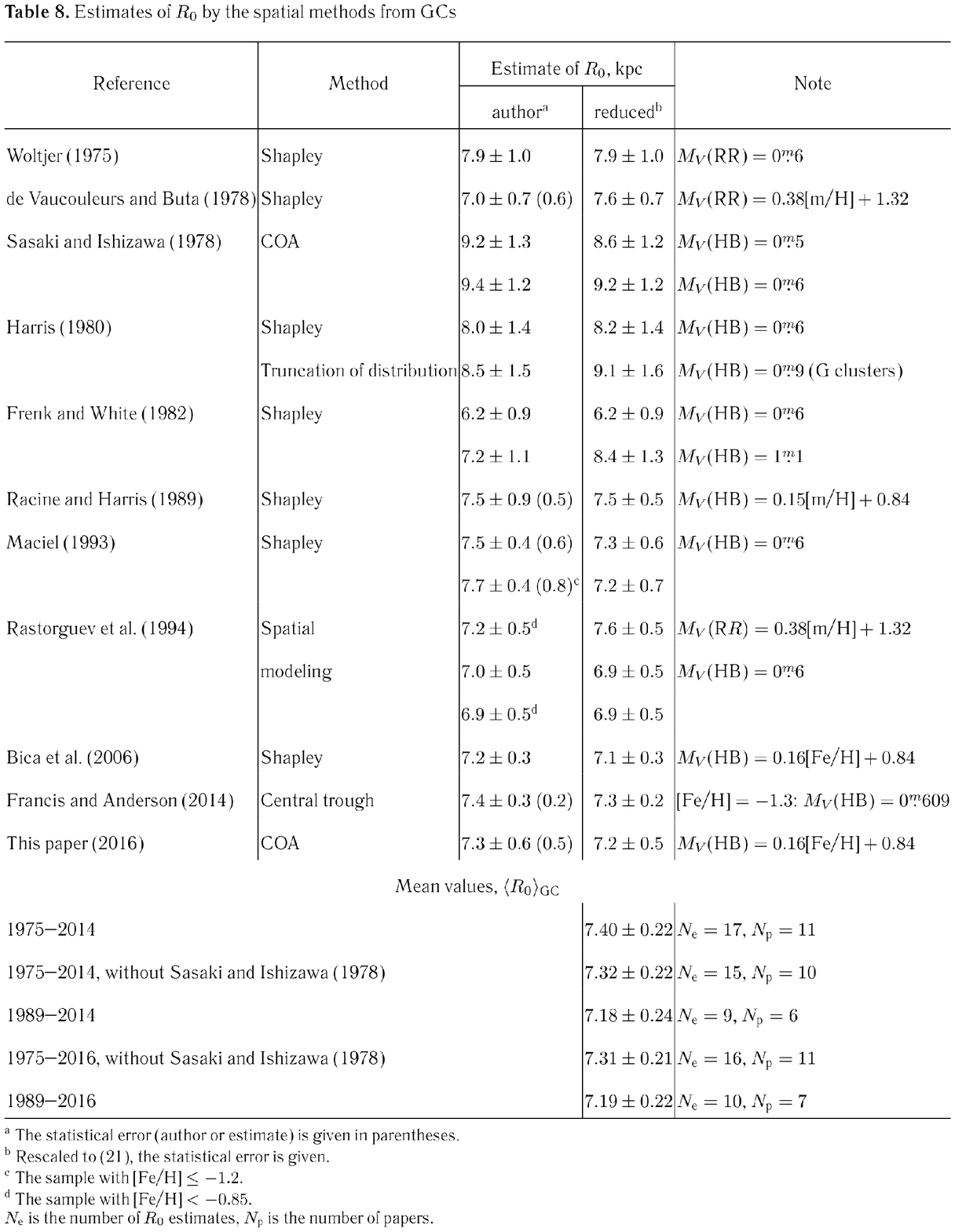}%
}%
%\vspace{-6em}
%\label{maxcone}
\end{figure}
%xxxxxxxxxxxxxxxxxxxxxxxxxxxxxxxxxxxxxxxxxxxxxxxxxxxxxxxxxxxxx

Thus, a random realization of the zone of avoidance
is unlikely. The fact that its axis will be orthogonal
to the Galactic plane by chance is even less likely
(Fig.~7).

Note that even the random nature of the zone of avoidance (which must not be ruled out
completely) does not abolish the very fact of its existence at the present epoch, because
the random errors in the distances to GCs at large $|b|$ are small (${\sim}0\m1\approx
5\%$, see H10) compared to the sizes of the identified axial voids, nor does it abolish
the fact that the position of the axis of this zone is close to the GC density maximum
(Fig.~7, Tables 3, 4).

If the axial zone of avoidance is nevertheless caused dynamically, then how stable is it
as a structure, given that some GCs can move in chaotic orbits (for example, NGC~6626; see
Casetti-Dinescu et al.\ 2013)? In recent years the proper motions have been measured for
many GCs, which has made it possible to calculate their orbits (see, e.g., Allen et al.
2006, 2008; Casetti-Dinescu et al.\ 2013). For example, in these three papers the
meridional orbits are provided for a total of 25 GCs. This gives some statistics for
answering the above question. During their orbital motion six of these GCs cross the
region of a formal COA with $\alpha_0=15^\circ$. However, all these GCs are relatively
close to the Galactic axis ($R_{\text{max}}=3$--8~ kpc) and enter into the COA region at
small distances from the Galactic plane: three GCs (NGC 6316, NGC 6528, and NGC 6626) at
$|Z|\la 1$~kpc (Allen et al.\ 2006; Casetti-Dinescu et al.\ 2013), two GCs (NGC 4833 and
NGC 6723) at $|Z|\la 2$~kpc (Allen et al.\ 2006, 2008), and only one GC (NGC 5968) at
$|Z|\la 4$~kpc (Allen et al.\ 2008). Besides, these GCs spend an insignificant fraction of
time in the COA, crossing it (often tangentially) near the pericenters of their orbits. In
any case, the outer COA regions (at least at $|Z|\ga 4$~kpc) seem stable structures from
this viewpoint; only the inner parts of the COA ($1\la |Z|\la 4$~kpc) can be periodically
``washed out.'' In the layer $|Z|\la 1$~kpc the COA is not revealed with confidence even
at the present epoch, in agreement with the statistics of orbits that independently
identifies this boundary. These results are largely explained by the fact that the
stochastization of orbits is enhanced by the bar and the spiral structure, which
predominantly affect, respectively, the GCs close to the Galactic center and the GCs with
a low energy of their vertical oscillations (see the above papers). In contrast, the COA
is revealed mainly by distant GCs at large $|Z|$. Note that stochastization does not
always increase the probability of GC entry into the COA region: for example, the orbit of
NGC 5968 in an axisymmetric potential crosses the COA by its loops more densely than in
the case where the bar and spirals are taken into account (Allen et al.\ 2008). Obviously,
this statistics of orbits, along with other arguments (see Subsection 5.2), stimulates the
parametric modeling methods in which the location of clusters in the formal zone of
avoidance is not ruled out completely, while the zone itself does not include the central
region of the Galaxy. 
\medskip

%\pagebreak[4]

%.............................................................
\subsection{The Quantity $R_0$}

We determined $R_0$ by modeling the distribution of Galactocentric latitudes for GCs by
two methods each of which has its advantages, difficulties, and weak points. The first
method (least-squares optimization, Subsection 5.1) is based on a sample of clusters that
in their $Z$ layers constrain best the position of the zone of avoidance. Other GCs,
including those at small $|Z|$ where the axial structure of avoidance is not revealed
unambiguously, are ignored. This makes the first method more refined in the sense of using
the COA as a structural feature to determine $R_0$ (here, no information about the number
density peak or the centroid of clusters is invoked). The absence of GCs close to the
Galactic plane in the final sample offers yet another advantage: the problems of selection
and extinction correction when determining the photometric distances are not significant
here. However, this method requires a preselection of the set of those GCs that outline
the axial structure using a special algorithm. At the same time, this method inherits the
errors of the operation of the selection algorithm, in particular, the fall of objects
with large deviations from the final axial direction into this set (Figs. 7, 8b) due to
the discreteness of the GC system. In addition, the limited number of GCs in the Galaxy
does not allow a denser partition into $Z$ layers to be performed and inevitably leads to
a small size of the sample of GCs outlining the axial zone of avoidance. The latter, in
turn, limits the statistical accuracy and robustness of the result (Table 5).

The second method of determining $R_0$ ($\chi^2$ minimization) uses not only the COA but
also, indirectly, the fact of GC concentration toward the Galactic center (the observed
distribution of latitudes $\varphi$ must be in best agreement with the model one at the
system's center even in the absence of COA). One advantage of the method is that it does
not require any a priori data on the zone of avoidance except the general form of its
model. Another advantage is the applicability of the method to a wide class of GC
subsamples, including the complete GC sample, i.e., the possibility of using samples of a
large size. This leads to statistically more accurate and robust $R_0$ estimates (Table 6)
than those yielded by the first method. One of the shortcomings of $\chi^2$ minimization
is that the objective function is nonsmooth, but this creates only computational
difficulties. The use of GCs near the Galactic plane can be a more serious shortcoming in
terms of systematic errors. Indeed, the observational incompleteness of GCs in the
Galactic disk and especially the presence of a dust bar (Schultheis et al.\ 2014), which
shields the GCs in the Galactic bar and behind it along the line of sight (Nikiforov and
Smirnova 2013), can noticeably shift the apparent symmetry center of the GC system to the
near side and, consequently, can underestimate $R_0$\,. As has been mentioned in the
Introduction, this is a potential source of systematic errors characteristic for the
spatial methods of determining $R_0$ from GCs.

A truncation of the GC distribution behind the Galactic center, $X > R_0$, in contrast
with the distribution at $X <R_0$ manifests itself for the layer $-0.58 \le Z < 1.0$~kpc
even in projection onto the $XZ$ plane (Fig.~7), with the truncation sharpness increasing
with decreasing $|Z|$. Under a noticeable influence of such selection on the
$\chi^2$-minimization result one might expect the $2\sigma$-interval estimates of $R_0$ to
be smaller than the formal ones, because the former are not directly associated with the
optimal position of the COA, while the latter are determined precisely by it. In reality,
the reverse is true: the $2\sigma$-interval estimates, on average, turn out to be even
slightly larger (by $0.14 \pm 0.10$~kpc) than the formal estimates (Subsection 5.3). The
exclusion of high-metallicity ($\text{Fe/H}]>-0.8$) GCs, for which the selection effect
must be stronger, barely changes the $R_0$ estimates (Table 6). Finally, the first method
of determining $R_0$ based on a sample of GCs among which there are no objects with
$|Z|\le 1.4$~kpc gives $R_0={7.29}^{+0.29}_{-0.28}$~kpc coincident with the result of the
second method, $\overline{R_0}=7.3\pm0.5$~kpc. All of this suggests that the observational
incompleteness of GCs does not affect strongly the $\chi^2$-minimization results. Note
that in both methods the minimum of the objective function establishing an optimal $R_0$
is identified unambiguously, in contrast to the method of maximizing the formal COA
(cf.~Fig.~1a with Figs. 8а and 9a).

As an estimate of $R_0$ from the COA in the distance scale (1) we take the result of the
second method, ${R_0}=7.3\pm0.5$~kpc, based on larger statistics and tested numerically.
The distance scale (1) was calibrated in H10 (the catalog by Harris (1996), the 2010
version) using the data on GCs in M 31, i.e., under the assumption of some distance to
this galaxy. This makes the calibration partly secondary. Let us rescale the results of
Subsection~5.3~to
\begin{equation}
    \label{H10calG}
    M_V(\text{HB})=0.165\,\text{[Fe/H]}+0.86,
\end{equation}
with the primary calibration obtained in H10 from the most direct distance measurements
{\em within our Galaxy\/} by comparing the scales (1) and (21) at the mean metallicity
$[\text{Fe/H}]=-1.3$ for all GCs and at the mean metallicity $[\text{Fe/H}]=-1.5$ for GCs
with $[\text{Fe/H}]<-0.8$. Then, $\overline{R_0(\chi^2_0)}=7.17$~kpc and
$\overline{R_0(2\sigma)}=7.31$. Averaging these estimates with weights of $(1.3)^2$ and $1$, we
finally obtain 
\begin{equation}
    \label{R0final}
    {R_0}=7.2\pm 0.5\bigr|_{\text{stat}} \pm
    0.3\bigr|_{\text{calib}}~\text{kpc}.
\end{equation}
Here, the second (systematic) error reflects
the $\pm 0\m1$ uncertainty in the calibration of the distance scale (21) (see H10).

The estimate (22) is small in comparison with the present-day means $\langle
R_0\rangle_{\text{best}}=(7.8$--$ 8.25) \pm (0.1$--$ 0.5)$~kpc (for references see
Section~1), but not in comparison with other $R_0$ determinations based on the spatial
distribution of GCs. Table 8 summarizes the estimates of this class published since 1975
and the averaged (see Nikiforov 2003, 2004) values, $\langle R_0\rangle_{\text{GC}}$ (GC
stands for globular clusters), for some of their subsets (the~systematic error of the
methods was taken to be $\sigma_{\text{meth}}=0\m2$). Our result agrees well with $\langle
R_0\rangle_{\text{GC}}$ obtained from previous estimates. Table 8 shows that beginning
from 1989 the $R_0$ estimates, being rescaled to (21), turn out to be close to one another
and they all do not exceed 7.6 kpc. The mean of them $\langle R_0\rangle_{\text{GC}}=7.18
\pm  0.24$~kpc is very close to our estimate (22). Averaging these estimates by taking
into account (22), i.e., over the more homogeneous group of 1989--2016 results, leads to
the current mean value of the spatial $R_0$ estimates from the data on GCs:
\begin{equation}
    \label{R0GC}
    \langle R_0\rangle_{\text{GC}}=7.19 \pm 0.2{\color{black}2}
    \bigr|_{\text{\color{black}stat, meth}} \pm 0.33\bigr|_{\text{\color{black}calib}}~\text{kpc}.
\end{equation}

Thus, our analysis of the spatial GC distribution leads to values of $R_0$ considerably
smaller than the bulk of the results obtained by other methods. The causes of this
discrepancy can be: (1) the underestimation of the present-day GC distance scale;
(2)~the~neglect of the selection effect and/or other systematic errors in the analysis of
the spatial GC distribution; and (3) the unrealistic assumption about a symmetry of the GC
system and/or a shift of the central spatial features of this system relative to the
Galactic ``centers'' determined in a different way. An attempt to explain the discrepancy
{\em only\/} by the first cause leads to the assumption that the distance scale of GCs
(and RR Lyrae variables ``akin'' to them in calibration) was underestimated by
$(10$--$13)\pm 5\%$, i.e., by $(0\m20$--$0\m26)\pm 0\m1$, which now seems not very
plausible [see the calibration in H10 and a summary of distance scales in Francis and
Anderson (2014)]. The second cause is more probable, but the closeness of the $R_0$
estimate fromthe COA in this paper to other recent spatial estimates from GCs argues
against a strong influence of the selection effect on $R_0$ in this class of methods. Note
that the disregarded systematics can also be in other methods, but it is unlikely to be
able to strongly shift (increase) the ``best'' mean $\langle R_0\rangle_{\text{best}}$ due
to the great variety of methods. In our view, the third (hypothetical) cause is not
improbable. Although the inner part of the Galaxy seems a well-established system where
the density peak of the visible matter may be deemed close to the barycenter (the center
of mass), this assumption almost certainly breaks down on scales of tens of kpc due to its
accretion and interaction with other galaxies of the Local group, which manifests itself,
for example, as an outer disk warp (see, e.g., Bland-Hawthorn and Gerhard 2016).
Therefore, the asymmetry of the GC system must not be ruled out completely, at least on
this scale. The difference between $\langle R_0\rangle_{\text{best}}$ and $\langle
R_0\rangle_{\text{GC}}$ may be due to a combination of all these causes. The question
remains~open.

Note that comparatively low values of $R_0$ have recently been obtained not only from an
analysis of the spatial GC distribution but also by some other methods, for example,
$R_0=7.52 \pm 0.10\bigr|_{\text{\color{black}stat}} \pm 0.35
\bigr|_{\text{\color{black}system}}$~kpc (Nishiyama et al.\ 2006) and $R_0=7.5
\pm  0.3$~kpc (Francis and Anderson 2014) from red clump stars in the bulge, $R_0=7.25 \pm 
0.32$~kpc (Bobylev 2013) from the kinematics of star-forming regions near the solar
circle, and $R_0=6.72 \pm  0.39 $~kpc (Branham 2014) from the kinematics of OB stars.
\medskip

%.............................................................
\subsection{The Parameter~$\varphi_0$}

As the final result for this parameter we take the estimate of
$\varphi_0=75\fdg0^{+4\fdg1}_{-2\fdg1}$ derived in Subsection 5.3. Note that it is very
close to $\varphi_0=75\fdg3$ obtained by a different method, from the void-forming GCs of
the axial zone of avoidance (the complete sample, see Subsection~5.1).

The error in the calibration of the GC distance scale does not distort the geometry of the
zone of avoidance and, consequently, does not lead to any systematic error in
the~$\varphi_0$ estimate. Since the COA is identified by GCs at large $|Z|$, the selection
effect for it is weak and can manifest itself only in a differential form: GCs in the far
part of the COA surface with respect to the Sun can be revealed with a lower probability
than those in the near part. In principle, this could lead to some overestimation of
$\varphi_0$ and $R_0$\,. However, the differential selection cannot be significant,
because although the differential effect grows with $|b|$, the selection itself decreases
sharply. Indeed, in the far (with respect to the COA) part of the Galaxy at $|Z|\ga 1$~kpc
many GCs have been detected at large distances (Figs.~5,~7).
\medskip

%.............................................................
\subsection{The Shape of the Zone of Avoidance}

The assumption about the existence of COA in the Galactic GC system arose from the fact
that a deficit of clusters toward the region along the Galactic axis was noticed in
projection onto the $XZ$ plane (Wright and Innanen 1972b). In Fig.~11a the GC distribution
in this projection based on the H10 data is compared with the COA contour for the
parameters~$\alpha_0=15\fdg 0$ and $R_0 = 7.3$~kpc found in this paper. Indeed, the
deficit of GCs in the axial biconical region is quite noticeable. However, this deficit
does not manifest itself clearly in projection onto the $YZ$  plane also parallel to the
COA axis (Fig.~11b). This is confirmed by the following statistics. The number of GCs
within the COA (i.e., being projected onto the COA) is $N_0 = 13$ for the $XZ$ plane and
$N_0 = 27$ for the $YZ$ plane. After the exclusion of clusters with $\text{[Fe/H]}>-0.8$,
these numbers are 12 and 23, respectively, which gives a density contrast of 1.9. In this
case, half of the GCs within the COA on the $XZ$ plane are very close to the formal COA
contour and largely form the apparent COA outline. If we exclude such clusters, i.e., take
$\alpha_0={\color{black}14\fdg 0}$, then $N_0^{XZ}=6$ and $N_0^{YZ}=21$ (for the  $XZ$ and
$YZ$ planes, respectively); the contrast is then 3.5. How probable is it to obtain such
numbers $N_0$ by chance?

The mean and variance of the number of clusters
visible within an axisymmetric COA with a half-angle
$\alpha_0$ for the distribution (14) are
\begin{equation}
    \label{N0}
    \mex N_0=Nn_0,\qquad \var N_0 = Nn_0(1-n_0), 
    \qquad n_0=\frac{2\alpha_0}{\pi\cos \alpha_0}-\frac{1}{\cos \alpha_0}+1,\nonumber
\end{equation}
where $N$ is the total size of the GC sample (see Appendix A3). At $N = 118$ (the sample
without GCs with $\text{[Fe/H]}>-0.8$) and $\alpha_0=14\fdg 0$ we obtain $\mex N_0=15.3\pm
3.{\color{black}7}$. The excess of $N_0^{YZ}=21$ above $\mex N_0$ can be explained by
chance: the probability $P(N_0\ge 21)={\color{black}8.1\%}$ (in accordance with the
binomial distribution) is not low. However, the underestimation of $N_0^{XZ}=6$ with
respect to $\mex N_0$ is more significant: $P(N_0\le 6)=4.1\times 10^{-3}$, i.e., the
hypothesis about an {\em axisymmetric\/} COA when projected onto the $XZ$ plane is
rejected with a probability of 99.59\%. The probability for such a COA to obtain the same
or larger difference for the two projections by chance approximately is $P(N_0^{XZ}\le
6)\cdot P(N_0^{YZ}\ge 21)=3.{\color{black}3}\times 10^{-4}$.

These results suggest that the axial zone of avoidance in the Galactic GC system is not
strictly axisymmetric but is elongated in a direction approximately orthogonal to the $X$
axis, i.e., the Galactic center--anticenter line. We apparently owe the very fact of the
detection of a zone of avoidance precisely to this combination of circumstances. First, if
the latter at the present epoch were not oriented (by chance) parallel to one of the
planes of the universally accepted Galactic cartesian coordinate system, then it would
probably be noticed much later (Fig.~11b illustrates how difficult it is to suspect its
presence for an ``unfortunate'' projection). Second, if the zone of avoidance were exactly
axisymmetric (conical), then the observed surface density contrast~$\widetilde{C}_0$ of
GCs visible within the COA with respect to the remaining part of the GC system would be
too weak and would not differ significantly from unity even in projection onto a plane
parallel to the COA axis: $\mex \widetilde{C}_0=0.80$, $\sigma(\widetilde{C}_0)= 0.21$ and
$\mex \widetilde{C}_0=0.82$, $\sigma(\widetilde{C}_0)= 0.22$ for $\alpha_0=15\fdg 0$ and
$14\fdg 0$, respectively; the means and standard errors of the~$\widetilde{C}_0$ estimate
at $N = 118$ are specified here [see Appendix A3, Eqs. (48) and (47)]. In contrast, when
projected onto a plane inclined to the COA axis, the density contrast will be even weaker,
because the central (denser) region of the GC system will occupy a larger fraction of the
COA area. Thus, the COA would not be noticeable in projection onto any plane and could be
detected only in a three-dimensional analysis. In reality, owing to the elongation of the
zone of avoidance, the observed contrast for the $XZ$ plane is much stronger and differs
significantly from unity: $\widetilde{C}_0=0.57\pm 0.17$ at $\alpha_0=15\fdg 0$ and
$\widetilde{C}_0=0.29\pm 0.12$ at $\alpha_0={\color{black}14\fdg 0}$; therefore, it was
noticed [here, we use Eqs. (44) and (46) from Appendix A3]. Note that the latter value
of~$\widetilde{C}_0$ also differs significantly from $\mex \widetilde{C}_0=0.82$. For the
$YZ$  plane the contrast is even reverse,  $\widetilde{C}_0=1.21\pm 0.29$ and $1.18\pm
0.28$ at $\alpha_0=15\fdg 0$ and ${\color{black}14\fdg0}$, respectively, but it differs
insignificantly from unity.

The subject matter being discussed is closely related
to the question about the detectability of similar
regions of avoidance in the GC systems of other
galaxies, because only an analysis of the cluster distribution
in the plane of the sky is accessible in this
case. We will assume that the region of avoidance is
revealed at a statistically significant level if
\begin{equation}
\label{C0:3sigma}
    \frac{\sigma(\widetilde{C}_0)}{1-\widetilde{C}_0}<\frac{1}{3}.
\end{equation}
Suppose that a galaxy is observed from an optimal
angle, i.e., edge-on. In the case of an axisymmetric
double COA, substituting Eq.~(47) and, to simplify
the estimation, Eq.~(43) for the asymptotic value~$C_0$
of the observed contrast~$\widetilde{C}_0$ into (24), we then obtain
a constraint on the number of GCs:
\begin{equation}
\label{N:3sigma}
    N> N_{3\sigma}= \frac{9 \cos^2 \alpha_0}{(1-\cos \alpha_0)^2}\cdot
	\frac{2\alpha_0+\pi\cos\alpha_0-\pi}{\pi-2\alpha_0}.
\end{equation}
The minimum number of GCs is $N_{3\sigma}=1.15\times 10^3$ at $\alpha_0=15\fdg 0$
($C_0=0.7956$) and $N_{3\sigma}=1.4{\color{black}3}\times 10^3$ at $\alpha_0={14\fdg 0}$
($C_0=0.8090$). Thus, provided that the axis of the COA lies in the plane of the sky, it
can be reliably detected only in rich GC systems that are rare among spiral galaxies; in
any case, the parent galaxy must be massive, with an absolute magnitude $M_V\la
-21{\color{black}\m}3$ (see, e.g., Figs. 5.5 and 5.6 in the book by Ashman and Zepf
(2008)).

If the zone of avoidance is nonaxisymmetric (elongated), then a large $N$ is not required
for its detection. Substituting Eqs. (43) and (46) into (24), we derive the constraint 
\begin{equation}
\label{N:3sigma_n0}
    N> N_{3\sigma}= \frac{9n_0}{1-n_0}\left(\frac{\pi-2\alpha_0}{2\alpha_0-\pi n_0}\right)^2.
\end{equation}
Taking the observed contrasts to be $C_0 = 0.57$ ($n_0=0.102$, $\alpha_0=15\fdg 0$) and
$C_0 = 0.29$ ($n_0=0.0508$, $\alpha_0={\color{black}14\fdg 0}$), we find $N_{3\sigma}=168$
and $3{\color{black}1}$, respectively. Thus, such zones of avoidance can also be
detected in principle in moderately populated GC systems, but this additionally requires
that the line of sight lie almost in the plane along which the zone is elongated, which
sharply reduces the detection probability. The latter also depends on the factors causing
the elongation.

The nonaxisymmetric structure of the zone of avoidance in the Galactic GC system may be
due to the influence of the Magellanic Clouds (MCs). This assumption is consistent with
some recent results. For example, there is evidence for the association of some of the
halo GCs with the Magellanic plane or the Vast Polar Structure (see, e.g., Pawlowski et
al.\ 2014). Yankelevich (2014) showed that Galactic GCs could even change the direction of
their rotation to the opposite (retrograde) one under the gravitational perturbation from
the MCs. Our assumption is also supported by the fact that in projection onto the $XZ$
plane the LMC is within the zone of avoidance, while the SMC is close to its contour
(Figs. 6, 7). The passage of the MCs through the southern part of the Galactic halo at the
present epoch may also explain the less regular shape of the southern cavity of the zone
of avoidance (cf. the same figures).

The possibility that the central trough in the GC distribution orthogonal to the~$X$~axis
that was identified by Francis and Anderson (2014, below referred to as FA14) is a
manifestation of the elongated zone of avoidance must not be ruled out. However, the
status of this FA14 result is not quite clear. The trough was identified not during the
spatial analysis but as one of the dips in the GC distribution in $X$ coordinate subjected
to Gaussian smoothing. This dip appears distinct and ``central'' only when using the
author's catalog of GC distances but not for the catalogs of H10 and Bica et al.\ (2006):
for H10 the dip is shallow and double, at $X_0\approx -0.2$  and $-1.0$~kpc ($X_0\equiv
X-7.4\text{~kpc}$); for the catalog by Bica et al.\ (2006) the dip is more pronounced but
lies at $X_0\approx -0.7$~kpc and then $R_0 = 6.7$~kpc (cf. Figs. 3 and 4 in FA14). The
distance scale adopted in FA14 differs from the distance scales of the two other catalogs
in that, being quadratic in metallicity, it assigns smaller distances to metal-rich GCs
(see Fig.~1 in FA14), which are located predominantly in the central Galactic region. This
displaces such GCs closer to the Sun, which in combination with the nondetectability of
GCs in the far part of the bar and behind the bar (Nikiforov and Smirnova 2013) can lead
to the ``trough effect.'' The extent to which the trough is attributable precisely to the
GCs close to the center is not pointed out directly in FA14 (the constraints on the
Galactocentric distance was imposed only from above, with the strongest one being 10 kpc),
but our assumption is supported by the fact that the central dip for the sample of
metal-poor GCs from the FA14 catalog is indistinct and almost merges with the neighboring
dip at $X_0\approx -0.8$~kpc (Fig.~3 in FA14). The zone of avoidance considered in this
paper is revealed, on the contrary, by GCs outside the Galactic bulge ($|Z|>1$~kpc).
Unfortunately, the question about the statistical significance of the central dip as a
structural feature for the constructed GC distribution in $X$ is not considered in FA14.
\medskip

%.............................................................
\subsection{Prospects for the Method}

In future, it would be desirable to use some modeling procedures that do not require any
obligatory constraints on the GC sample. The $\chi^2$ minimization belongs to these
procedures, but the objective function for it turns out to be nonsmooth. This shortcoming
can be overcome by passing to the maximum likelihood method (MLM). Attempts to apply it
for model (16) (Nikiforov and Agladze 2013) confirm the existence of a region of
avoidance: $\varphi_0=77\fdg4^{+2\fdg5}_{-1\fdg7}$ for the complete GC sample and
$\varphi_0=76\fdg8^{+3\fdg3}_{-1\fdg0}$ for the sample with $[\text{Fe/H}]<-0.8$. In this
case, however, the $R_0$ estimates turned out to be generally shifted (6.01--7.02 kpc) due
to the presence of a high hump in the profile of the logarithmic likelihood
function~${\cal L}_{\text{m}}(R_0)\equiv -\ln L_{\text{m}}(R_0)$ at slightly larger $R_0$
(i.e., in the region of the highest GC density). The~${\cal L}_{\text{m}}(R_0)$ profile
leftward of the point estimate does not reach even the $2\sigma$ level (in contrast to
the~~$\chi^2_{\text{m}}(R_0)$ profile). Studies have shown that the high sensitivity of
the likelihood function to the location of central GCs (with small $|Z|$) at $|\varphi|$
very close to $90^\circ$ is responsible for the first effect. In this case, at the
corresponding $R_0$ the probabilities to find such clusters according to model (16) are
low at any values of other parameters. This gives the spikes in ${\cal
L}_{\text{m}}(R_0)$. When using the $\chi^2$ minimization, nothing of the kind occurs due
to the smoothing: at~$\varphi_0$ slightly larger than $80^\circ$ the terms
$\frac{({\nu}_i-N{p}_i)^2}{N{p}_i}$ for such GCs are not too large. The second effect is
caused by the systematic drop in ${\cal L}_{\text{m}}(R_0)$ as $R_0$ recedes from the true
value due to the appearance of an increasingly large number of GCs at small $|\varphi|$,
for which the occurrence probability according to (16) is maximal. The~$\chi^2$ statistics
tests the shape of the observed $|\varphi|$ distribution, which is strongly deformed as
$R_0$ recedes from the true one and poorly corresponds to model (16) at any values of
other parameters; therefore, $\chi^2_{\text{m}}(R_0)$ does not fall off to the edges of
the $R_0$ interval considered (Fig.~9a). Thus, the $\chi^2$ minimization turned out to be
a more efficient estimator for model (16). For these reasons, we do not use the results
obtained by the MLM in this paper.

However, the introduction of a more adequate model for the bulge component of the GC
system can make the MLM efficient. This was highly desirable for the solution of our
problem in a more general form.

The sharp truncations of distribution (16) at $\varphi=|\varphi_0|$ lead to the
nonsmoothness of any objective function. The introduction of blurred truncation
boundaries, which may also turn out to be more physical, would allow the problem to be
solved.

Using a smooth objective function will allow more difficult problems (under more general
assumptions), including the modeling of the zone of avoidance when abandoning the
assumption of an axial symmetry, to be solved. Note that allowance for the elongation of
the zone of avoidance will not shift the region of intersection of the zone projection
onto the $XY$ plane with the $X$ axis, i.e., will not affect significantly the $R_0$
estimate.
\medskip

%***************************************************************
\section{CONCLUSIONS}

The possibility of the presence of a double cone of avoidance (COA) in the globular
cluster (GC) system of the Galaxy oriented along the Galactic rotation axis was considered
previously in the literature. This direction was not developed further, but the only use
of this structural feature to determine the distance to the Galactic center ($R_0$) in
Sasaki and Ishizawa (1978) continued to be taken into account in the context of the $R_0$
problem. The main goal of this paper is to check whether an axial zone of avoidance exists
in the~GC system and to perform its parametrization that includes the new determination of
$R_0$ as its parameter from currently available data (the catalog by Harris (1996), the
2010 version).

We showed that applying the method of maximizing the formal COA in Sasaki and Ishizawa
(1978) generally led to an incorrect result because of the off-axis COAs due to the
discreteness of the GC system. The presence of several approximately equal formal COAs
revealed by the maximization method does not allow the existence of an {\em axial\/} zone
of avoidance to be judged with confidence.

To check whether this zone was real, we represented the voids in the GC system by a set of
largest-radius meridional cylindrical voids. As a result of our separate and joint
analyses of the northern and southern voids, we managed to identify an ordered axial zone
of avoidance and constructed the maps of its section by a meridional plane. The results
obtained argue for the existence of a zone of avoidance for GCs along the Galactic axis
(outside the small central region) that is close in shape to a double cone. The northern
and southern COA cavities manifest themselves independently and with similar parame- ters.
The position of the COA vertex is close to the position of the GC number density maximum.

Modeling the distribution of Galactocentric latitudes for GCs allows the optimal position
of the axial COA to be unambiguously identified and leads to the following estimates of
the parameters: $\varphi_0=75\fdg0^{+4\fdg1}_{-2\fdg1}$, i.e., the COA half-angle
$\alpha_0={15\fdg 0}^{+2\fdg1}_{-4\fdg1}$, ${R_0}=7.3\pm0.5$~kpc. The southern COA cavity
can be wider ($\alpha_0\approx 18\deg$) than the northern one ($\alpha_0\approx 12\deg$),
but this difference is not significant. Comparison of the modeling results for various
techniques and GC samples shows that the observational incompleteness of GCs affects
insignificantly the $R_0$ estimate. A correction to the H10 calibration of the GC distance
scale based on Galactic objects leads to an estimate of ${R_0}=7.2\pm
0.5\bigr|_{\text{stat}} \pm 0.3\bigr|_{\text{calib}}~\text{kpc}$. The latter is, on
average, smaller than the present-day $R_0$ estimates but agrees well with $R_0$ deduced
by analyzing the spatial GC distribution ($\langle R_0\rangle_{\text{GC}}=7.19 \pm 0.22
\bigr|_{\text{stat, meth}} \pm 0.33\bigr|_{\text{calib}}$~kpc).

As the maps of the zone of avoidance suggest, its southern cavity has a less regular shape
than the northern one, which may be due to the influence of the Magellanic Clouds.
Comparison of the GC distributions on two meridional planes revealed evidence for the
elongation of the zone of avoidance in a direction orthogonal to the center--anticenter
direction. This nonaxisymmetric structure may be caused by the same factor.

Our numerical experiments showed that in the absence of a region of avoidance the
probability to obtain the same COA as that from real data or wider and, at the same time,
$R_0$ estimates as close as those from real data or closer to the mean $\langle
R_0\rangle_{\text{GC}}$ does not exceed 2\%.

In external galaxies even when they are seen edge-on an axisymmetric COA can be detected
only for rich GC systems ($N_{\text{GC}}>10^3$). Nonaxisymmetric zones of avoidance can
also be revealed in principle in moderately populated GC systems, but only at a (lucky)
orientation of the elongation of the zone along the line of sight.
\medskip

%---------------------------------------------------------------

\section*{\sl\hfill APPENDIX}
\subsection*{A1. Distribution of the Absolute Value of the
Galactocentric Latitude}

Obviously, the probability density of the random variable $\Phi$, the absolute values of
the Galactocentric latitude $\varphi$, for a spherically symmetric GC distribution without
any COA differs from the law (15) only by the normalization: 
\begin{equation}
\label{absphi}
f\left({|\varphi|}\right)=\cos{\varphi}. %\nonumber
\end{equation}
The mean of $\Phi$ is then 
\begin{equation}
    \mex \Phi=\int\limits^{\pi/2}_{0}{\varphi\cos{\varphi}\,d\varphi}=\frac{\pi}{2}-1
    \approx 0.5708 \approx 32\fdg 70.
\end{equation}
\medskip

\subsection*{A2. Distribution of the Largest of the Two Absolute
Values\\ of the Galactocentric Latitudes}

Consider the random variable~$\Phi_{\text{m}}=\max (\Phi_1,\Phi_2)$, where $\Phi_1$ and
$\Phi_2$ are the arguments of the pair of members (the absolute values of the latitudes
for the pair of globular clusters) extracted at random from a population with the argument
$\Phi$ obeying the law (27). Let us find the probability density $f(\varphi_{\text{m}})$
of the random variable $\Phi_{\text{m}}$\,. (The quantities $\varphi_{\text{m}}$ defined
for the axial voids in Subsections 4.2 and 4.3 may be considered as the values of the
random variable $\Phi_{\text{m}}$ in the absence of COA.)

Since the order in the pair of members is unimportant, we will assign the indices in each
pair of $\phi_1$  and $\phi_2$ values of the random variables $\Phi_1$ and $\Phi_2$, for
example, according to the rule $\phi_1\geq\phi_2$\,. Then, $\varphi_{\text{m}}=\phi_1$\,.
In view of (27), the probability that $\Phi_{\text{m}}$ will fall into the interval
$[\varphi_{\text{m}},\varphi_{\text{m}}+d\varphi_{\text{m}}]$ is defined as 
\begin{equation}
\label{prob_phi1}
\begin{split}
    P(\varphi_{\text{m}}\leq \Phi_{\text{m}} <\varphi_{\text{m}}+d\varphi_{\text{m}})&=
    P(\phi_1\leq \Phi_{\text{m}} <\phi_1+d\phi_1)=\\
    &=P(\phi_1\leq \Phi_1 <\phi_1+d\phi_1)\cdot P(\Phi_2\leq \phi_1)=\\
    &=c'_1 \cos \phi_1 \int\limits^{\phi_1}_0\cos \phi\,d\phi=c_1 \sin 2\phi_1=
    c_1 \sin 2\varphi_{\text{m}}\,,
\end{split}
\end{equation}
where $c'_1$  and $c_1$ are the variants of the normalization constants.
The choice of a different rule ($\phi_2\geq\phi_1$) and the permutation of indices 1 and 2 in
(29) do not change the result. Given the normalization
\begin{equation}
% \label{jjj}
1=c_1\int\limits^{\pi/2}_{0}{\sin 2\varphi_{\text{m}}\,d\varphi_{\text{m}}}=c_1\,,\nonumber
\end{equation}
we finally obtain 
\begin{equation}
%\label{absphi}
f(\varphi_{\text{m}})=\sin {2\varphi_{\text{m}}}\,. %\nonumber
\end{equation}

Then, the mean of $\Phi_{\text{m}}$ is
\begin{equation}
    \mex\Phi_{\text{m}}=
    \int\limits^{\pi/2}_{0}{\varphi_{\text{m}}\sin{2\varphi_{\text{m}}}\,d\varphi_{\text{m}}}=
    \frac{\pi}{4}\,,
\end{equation}
the variance of $\Phi_{\text{m}}$ is
\begin{equation}
    \var\Phi_{\text{m}}=
    \int\limits^{\pi/2}_{0}{\left(\varphi_{\text{m}}-\frac{\pi}{4}\right)^2
	\sin{2\varphi_{\text{m}}}\,d\varphi_{\text{m}}}=
    \frac{\pi^2-8}{16},
\end{equation}
and the standard deviation is
\begin{equation}
    (\var\Phi_{\text{m}})^{1/2}=\frac{\sqrt{\pi^2-8}}{4}\approx 0.3418\approx 19\fdg59.
\end{equation}
\medskip

\subsection*{A3. Surface Density Contrast between Globular
Clusters Inside and Outside the Cone of~Avoidance
when Observed in Projection onto a Plane Parallel
to the Cone Axis}

In a spherically symmetric GC system with a double axial COA, i.e., with the distribution
of Galactocentric latitudes (14), the number of GCs seen {\em in projection\/} onto the
COA, $N_0$, does not depend on the radial density law ${\mathfrak f}(R_\text{g})$ if the
plane of projection is parallel to the COA axis. Indeed, in this case, the radius drawn
from the system's center cannot cross the boundary of the region of space being projected
onto the COA, because two planes passing through the center form this boundary; the radius
can only lie at the boundary. $N_0$ is then completely determined by the total number of
GCs in the system, $N$, and the fraction, $n_0$, of the system's volume being projected
onto the COA, $V_0$\,:
\begin{equation}
    N_0=Nn_0\,,\qquad n_0=\frac{V_0}{V},
\end{equation}
where $V$ is the volume of the sphere minus the double cone. If the opening angle of the
latter is $\alpha_0$\,, then
\begin{equation}
    V=\frac{4}{3}\pi a^3 \cos\alpha_0\,,
\end{equation}
where $a$ is the radius of the GC system.

The volume $V_0$ is the difference between the volume of the double azimuthal sector of
the sphere
\begin{equation}
    V_\text{as}=\frac{4}{3}\pi a^3\cdot \frac{4\alpha_0}{2\pi}=\frac{8}{3}a^3 \alpha_0
\end{equation}
and the volume of the double spherical
sector
\begin{equation}
    V_{\text{ss}}=2\int^a_0\!\int^{2\pi}_0\!\!\!\int^{\pi/2}_{\varphi_0} 
	R^2_{\text{g}} \cos \varphi \,dR_{\text{g}}\,d\theta\,d\varphi =
    \frac{4}{3}\pi a^3(1-\cos\alpha_0),
\end{equation}
where $\varphi_0=\pi/2-\alpha_0$\,. Then,
\begin{gather}
    V_0=\frac{4}{3}a^3(2\alpha_0 +\pi\cos\alpha_0-\pi),\\
    n_0=\frac{2\alpha_0}{\pi\cos \alpha_0}-\frac{1}{\cos \alpha_0}+1.
    \label{n0}
\end{gather}
Hence the fraction of the system's volume that is not
projected onto the COA is
\begin{equation}
\label{n1}
    n_1=1-n_0=\frac{1}{\cos \alpha_0}-\frac{2\alpha_0}{\pi\cos \alpha_0}.
\end{equation}

Given $n_0$ and $n_1$, it is easy to find the GC surface
density inside the COA
\begin{equation}
    \mu_0=\frac{N_0}{S_0}=\frac{Nn_0}{\pi a^2}\cdot \frac{2\pi}{4\alpha_0}=
	\frac{Nn_0}{2\alpha_0 a^2}
\end{equation}
and the surface density outside the COA
\begin{equation}
    \mu_1=\frac{Nn_1}{S_1}=\frac{Nn_1}{\pi a^2} \frac{2\pi}{2\pi-4\alpha_0}=
	\frac{Nn_1}{a^2(\pi-2\alpha_0)},
\end{equation}
where $S_0$ and $S_1$ are the areas of the projection of
the double spherical sector and the projection of the
part of the GC system outside the COA, respectively.
Hence, given (39) and (40), we obtain a theoretical
value of the surface density contrast inside the COA
with respect to the remaining part of the GC system:
\begin{equation}
\label{C0mex}
    C_0=\frac{\mu_0}{\mu_1}=\frac{n_0}{n_1}\frac{\pi-2\alpha_0}{2\alpha_0}=
	1-\frac{\pi}{2\alpha_0}\,(1-\cos \alpha_0).
\end{equation}

Counting the GCs that fell within the COA gives
an estimate of the contrast from observations:
\begin{equation}
\label{C0est}
    \widetilde{C}_0=\frac{N_0}{N-N_0}\frac{\pi-2\alpha_0}{2\alpha_0}.
\end{equation}
Since $N_0$ as a random variable obeys a binomial distribution,
\begin{equation}
\label{N0:mex_var}
    \mex N_0= N n_0,\qquad \var N_0 = Nn_0(1-n_0).
\end{equation}
Using (45), it is easy to find the statistical error of the~$\widetilde{C}_0$
estimate from the error propagation formula
\begin{equation}
\label{C0:var_n0}
    \sigma(\widetilde{C}_0)=\frac{\sqrt{n_0}\,(\pi/2\alpha_0-1)}{\sqrt{N}(1-n_0)^{3/2}}.
\end{equation}
In general, we may adopt $n_0=N_0/N$ to estimate the uncertainty in the observed contrast.
In the case of COA, substituting Eq.~(39) for $n_0$, we obtain
\begin{equation}
\label{C0:var_alpha0}
    \sigma(\widetilde{C}_0)=\frac{\pi\cos \alpha_0}{2\alpha_0\sqrt{N}}
	\sqrt{\frac{2\alpha_0+\pi\cos\alpha_0-\pi}{\pi-2\alpha_0}}.
\end{equation}
For the same case, given (43)--(45), we find the mean of the observed
contrast:
\begin{equation}
\label{C0est:mex}
    \mex \widetilde{C}_0(N_0)\approx \widetilde{C}_0(\mex N_0)+
	\frac{1}{2}\widetilde{C}''_0(\mex N_0)\var N_0= 
    C_0\left[1+\frac{1}{N(1-n_0)}\right],
\end{equation}
i.e., $C_0=\lim\limits_{N\to\infty}\mex \widetilde{C}_0$\,.

%***************************************************************
\section*{ACKNOWLEDGMENTS}

We are grateful to V.A.~Marsakov and other (anonymous) referees for their useful remarks.
We also thank K.E.~Prokhorov and M.S.~Davydenkova for processing the early versions of the
data catalogs and A.V.~Veselova for performing auxiliary calculations.

This work was supported by the St.~Petersburg State University (grant no.\ 6.37.341.2015).

%***************************************************************
\section*{REFERENCES}

\begin{enumerate}

\item
T. A. Agekyan, {\em Probability Theory for Astronomers
and Physicists\/} (Nauka, Moscow, 1974) [in Russian].

\item
C.~Allen, E. Moreno, and B. Pichardo, Astrophys.\ J.
{\bf 652\/}, 1150 (2006).

\item
C. Allen, E. Moreno, and B. Pichardo, Astrophys.\ J.
{\bf 674\/}, 237 (2008).

\item
K. M. Ashman and S. E. Zepf, {\em Globular Cluster
Systems\/} (Cambridge Univ. Press, Cambridge, UK,
2008).

\item
B. S. Avedisova, Astron.\ Rep.\ {\bf 49\/}, 435 (2005).

\item
E. Bica, C. Bonatto, B. Barbuy, and S. Ortolani,
Astron.\ Astrophys.\ {\bf 450\/}, 105 (2006).

\item
J. Bland-Hawthorn and O. Gerhard, Ann.\ Rev.\ Astron.\ 
Astrophys.\ {\bf 54\/}, 529 (2016).

\item
V. V. Bobylev, Astron.\ Lett.\ {\bf 39\/}, 95 (2013).

\item
T. V. Borkova and V. A. Marsakov, Astron.\ Rep.\ {\bf 44\/},
665 (2000).

\item
R. L. Branham, Jr., Astrophys.\ Space.\ Sci.\ {\bf 353\/}, 179
(2014).

\item
D. I. Casetti-Dinescu, T. M. Girard, L. J\'{\i}lkov\'{a},
W. F. van Altena, F. Podest\'{a} , and C. E. L\'opez, Astron.
J. {\bf 146\/}, 33 (2013).

\item
T. Foster and B. Cooper, ASP Conf.\ Ser.\ {\bf 438\/}, 16
(2010).

\item
C. Francis and E. Anderson, Mon.\ Not.\ R.\ Astron.
Soc.\ {\bf 441\/}, 1105 (2014).

\item
C. S. Frenk and S. D. M. White, Mon.\ Not.\ R.\ Astron.
Soc.\ {\bf 198\/}, 173 (1982).

\item
R. Genzel, F. Eisenhauer, and S. Gillessen, Rev.\ Mod.\ 
Phys.\ {\bf 82\/}, 3121 (2010).

\item
W. E. Harris, in {\em Star Clusters, Proceedings of the
85th IAU Symposium, Victoria, BC, Canada, August
27--30, 1979\/}, Ed.\ by J. E. Hesser (Reidel, Dordrecht,
1980), p.~81.

\item
W. E. Harris, Astron.\ J. {\bf 112\/}, 1487 (1996);
arXiv:1012.3224 (2010).

\item
W. E. Harris, {\em Saas-Fee Advanced Courses\/}, Ed.\ by
L. Labhardt and B. Binggeli (Springer, Berlin, 2001),
Vol.~28, p.~223.

\item
F. J. Kerr and D. Lynden-Bell, Mon.\ Not.\ R.\ Astron.\ 
Soc.\ {\bf 221\/}, 1023 (1986).

\item
R. C. Kraan-Korteweg and O. Lahav, Astron.\ Astrophys.\ 
Rev.\ {\bf 10\/}, 211 (2000).

\item
A. V. Loktin and V. A. Marsakov, {\em Lectures on Stellar Astronomy\/} (Yuzh.\ Fed.\ Univ.,
Rostov-on-Don, 2009) [in Russian]. http://www.astronet.ru/db/msg/1245721/ \mbox{index}.html

\item
W. J. Maciel, Astrophys.\ Space.\ Sci.\ {\bf 206\/}, 285 (1993).

\item
I. I. Nikiforov, Astrophysics {\bf 42\/}, 300 (1999).

\item
I. I. Nikiforov, Cand.\ Sci.\ (Phys. Math.) Dissertation
(SPb.\ State Univ., St.\ Petersburg, 2003).
http://www.astro.spbu.ru/?q=nii

\item
I. I. Nikiforov, ASP Conf.\ Ser.\ {\bf 316\/}, 199 (2004).

\item
I. I. Nikiforov, Astron.\ Astrophys.\ Trans. {\bf 27\/}, 537
(2012).

\item
I. I. Nikiforov and E. V. Agladze, Izv. GAO {\bf 220\/}, 429
(2013).

\item
I. I. Nikiforov and E. E. Kazakevich, Izv. GAO {\bf 219\/}, 4,
245 (2009).

\item
I. I. Nikiforov and O. V. Smirnova, Astron.\ Nachr.
{\bf 334\/}, 749 (2013).

\item
S. Nishiyama, T. Nagata, Sh. Sato, D. Kato, T. Nagayama,
N. Kusakabe, N. Matsunaga, T. Naoi, et al.,
Astrophys.\ J. {\bf 647\/}, 1093 (2006).

\item
J. H. Oort, {\em Stars and Stellar Systems\/}, Vol.~5:
{\em Galactic Structure\/}, Ed.\ by A. Blaauw and
M. Schmidt (Univ.\ Chicago Press, Chicago, London,
1965), p.~455.

\item
M. Pawlowski, B. Famaey, H. Jerjen, D. Merritt,
P. Kroupa, J. Dabringhausen, F.~L\"{u}ghausen,
D. A. Forbes, et al., Mon.\ Not.\ R.\ Astron.\ Soc.\ {\bf 442\/},
2362 (2014).

\item
W. H. Press, B. P. Flannery, S. A. Teukolsky, and
W. T. Vetterling, {\em Numerical Recipes in C: The Art
of Scientific Computing\/}, 2nd ed.\ (Cambridge Univ.\ 
Press, Cambridge, 1997).

\item
R. Racine and W. E. Harris, Astron.\ J. {\bf 98\/}, 1609
(1989).

\item
A. S. Rastorguev, E. D. Pavlovskaya, O. V. Durlevich,
and A. A. Filippova, Astron.\ Lett.\ {\bf 20\/}, 591 (1994).

\item
M. J. Reid, in {\em The Center of the Galaxy, Proceedings
of the 136th IAU Sympoisum, Los Angeles,
USA, July 25--29, 1988\/}, Ed.\ by M. Morris (Kluwer
Academic, Dordrecht, 1989), p.~37.

\item
M. J. Reid, Ann.\ Rev.\ Astron.\ Astrophys.\ {\bf 31\/}, 345
(1993).

\item
T. Sasaki and T. Ishizawa, Astron.\ Astrophys.\ {\bf 69\/}, 381
(1978).

\item
M. Schultheis, B. Q. Chen, B. W. Jiang, O. A. Gonzalez,
R. Enokiya, Y. Fukui, K.~Torii, M. Rejkuba, et
al., Astron.\ Astrophys.\ {\bf 566\/}, A120 (2014).

\item
H. Shapley, Astrophys.\ J. {\bf 48\/}, 154 (1918).

\item
V. G. Surdin, Astron.\ Astrophys.\ Trans. {\bf 18\/}, 367
(1999).

\item
A. A. Sveshnikov, {\em Collection of Problems on Probability
Theory, Mathematical Statistics and Theory
of Random Functions\/} (Lan', St.~Petersburg, 2008),
p.~295 [in Russian].

\item
G. de Vaucouleurs, Astrophys.\ J. {\bf 268\/}, 451 (1983).

\item
G. de Vaucouleurs and R. Buta, Astron.\ J. {\bf 83\/}, 1383
(1978).

\item
L. Woltjer, Astron.\ Astrophys.\ {\bf 42\/}, 109 (1975).

\item
A. E. Wright and K. A. Innanen, Astron.\ Astrophys.
{\bf 21\/}, 151 (1972a).

\item
A. E. Wright and K. A. Innanen, Bull.\ Am.\ Astron.\ 
Soc.\ 4, {\bf 267\/} (1972b).

\item
V. Yankelevich, Astron.\ Astrophys.\ Trans. {\bf 28\/}, 347
(2014).

\end{enumerate}

\end{document}